# Electrochemical $CO_2$ capture with pH-independent redox chemistry

Sang Cheol Kim[1,‡], Marco Gigantino[2,‡], John Holoubek[3], Jesse E. Matthews[2,4], Junjie Chen[2,4], Yaereen Dho[5], Thomas F. Jaramillo[2,4,6], Yi Cui[3,6,7], Arun Majumdar[6,8], Yan-Kai Tzeng[9], Steven Chu[1,6,10*]

[1]Department of Physics, Stanford University, Stanford, California 94305, USA.

[2]Department of Chemical Engineering, Stanford University, Stanford, California 94305, USA.

[3]Department of Materials Science and Engineering, Stanford University, Stanford, California 94305, USA.

[4]SUNCAT Center for Interface Science and Catalysis, SLAC National Accelerator Laboratory, 2575 Sand Hill Road, Menlo Park, CA 94025, USA.

[5]Department of Chemistry, Stanford University, Stanford, California 94305, USA.

[6]Department of Energy Science and Engineering, Stanford University, Stanford, California 94305, USA.

[7]Stanford Institute for Materials and Energy Sciences, SLAC National Accelerator Laboratory, 2575 Sand Hill Road, Menlo Park, California 94025, USA.

[8]Department of Mechanical Engineering, Stanford University, Stanford, California 94305, USA.

[9]SLAC-Stanford Battery Center, SLAC National Accelerator Laboratory, 2575 Sand Hill Road, Menlo Park, California 94025, USA.

[10]Department of Molecular and Cellular Physiology, Stanford University, Stanford, CA 94305, USA.

‡ These authors contributed equally

Correspondence to: S. C. schu@stanford.edu

**Capture of anthropogenic $CO_2$ is critical for mitigating climate change, and reducing the energy cost is essential for wide-scale deployment[1–3]. Solubility of inorganic carbon in aqueous solutions depends on the pH, and electrochemical modulation of the pH has been investigated as a means of $CO_2$ capture and release[4–10]. However, reported methods incur unavoidable energy costs due to thermodynamic penalties. In this study, we introduce a pH-independent redox chemistry that greatly lowers the thermodynamic energy costs by changing the pH without directly changing the [$H^+$]. We show that the redox reaction of TEMPO molecules modulates the pH for capture and release of $CO_2$ in a flow cell with an energy cost as low as 2.6 kJ/mol of $CO_2$ corresponding to 0.027 eV/molecule. A molecular model, supported by MD and DFT simulations, is proposed of how the pH is decreased by 7.6 while largely avoiding the entropic energy cost associated with increasing the [$H^+$]. We believe that this work showcases the potential of pH-independent redox chemistries for practical and cost-effective $CO_2$ capture.**

**Introduction**

Capturing anthropogenic carbon dioxide ($CO_2$) will be a key strategy in reducing greenhouse gas emissions and meeting climate goals[1,2]. However, the high energy cost of current capture technologies based on chemical binding and release remains an obstacle to commercial deployment. Most capture and release methods require more than 100 kJ/mol[3], due to the irreversible energy loss in chemical bond forming/breaking of $CO_2$ bound to an absorbent and heating of inactive components such as the aqueous solvent. By comparison, the minimum Gibbs free energy $\Delta G$ needed to increase the partial pressure of $CO_2$ is determined by the change in entropy, given by $\Delta G = RT \ln\left(\frac{P_2}{P_1}\right)$, where $P_i, P_f$ are the initial capture and final release pressures of $CO_2$, respectively. Concentrating from a 20% $CO_2$ stream at a point source to 1 atm of $CO_2$ requires 4.0 kJ/mol while direct air capture starting from 400 ppm requires 19.4 kJ/mol (See Supplementary Note 1 and Supplementary Fig. 1 for calculations). Emerging strategies such as porous materials improve energy efficiencies[11–13], but the parasitic energy loss to heating inactive components remains difficult to avoid for thermochemical methods.

To circumvent such limitations, other methods have recently garnered attention, including electrochemical carbon capture[3,14–16]. Electrochemical activation can more specifically target active materials, mitigating energy losses. Several different electrochemical carbon capture approaches have emerged, including pH swing[4–7], redox-active sorbents[17–19], and electrochemically-mediated amine regeneration (EMAR)[20,21]. Among these approaches, the pH swing mechanism leverages the pH dependence of the solubility of inorganic carbon, including carbonates, bicarbonates and carbonic acid[15]. At high pH, the high solubility and reaction rates of inorganic carbon enable efficient $CO_2$ capture, while $CO_2$ is rapidly released at low pH. Thus, electrochemical modulation of the pH is an effective method of capturing and releasing $CO_2$[14,15,22].

Several different methods have been explored for electrochemical pH swing. Bipolar membrane electrodialysis (BPMED) dissociates water into $OH^-$ and $H^+$ that migrate to opposite sides of an electrochemical cell, creating a pH gradient[6,15]. Proton-coupled electron transfer (PCET) utilizes redox-active molecules that uptake or release $H^+$ upon reduction or oxidation to modulate the pH[5,8,9,23]. Another method is gas ($H_2$, $O_2$, etc.) looping[4,10]. For example, by coupling oxygen evolution and oxygen reduction reactions, a pH gradient can be created[4]. A fundamental property shared across these pH swing mechanisms is that $H^+$ or $OH^-$ is directly involved in the redox reaction. In all of the methods referenced above, the reduced molecule releases a proton (or consumes a hydroxide ion) upon oxidation. Because $H^+$ or $OH^-$ is directly involved as reactant or product, the reaction thermodynamics is dependent on the activity of these species. The Nernst Equation that determines the electrochemical energy needed to overcome the entropic penalty of directly modulating the concentration of $H^+$ is fully discussed in Supplementary Note 2 of the Supplementary Information. From these fundamental considerations, the *minimum* voltage needed to change the *pH* by changing the proton concentration $[H^+]$ is $-0.059$ V for each $\Delta pH = 1$.

In this work, we present a pH-independent redox chemistry for energy efficient $CO_2$ capture using TEMPO ((2,2,6,6-Tetramethylpiperidin-1-yl)oxyl), a redox-active stable free radical. Our method is based on modulating the pH by changing the activity coefficient $\gamma$ that appears in the definition of $pH \equiv -\log_{10}(\gamma[H^+])$. By circumventing direct modulation of $H^+$ concentration, TEMPO redox reaction is not constrained by the Nernstian energy penalty and the energy input is significantly lowered. In a flow cell system shown in schematically Fig. 1, , we show that an *operando* cell voltage of 0.022 V is sufficient to swing the pH by $\Delta pH \sim 7.6$, well below the minimum energy penalty of ~ 0.45 V required if the proton concentration [$H^+$] was changed.is . In the limit of low reaction rates, if the redox reaction is independent of $pH$, then the energy input of the oxidation/reduction cycle used to modulate the pH approaches zero.

Molecular dynamics (MD) and density functional theory (DFT) simulations show that electrochemically oxidized TEMPO acts as a charged Lewis acid that accepts electrons and polarizes surrounding water molecules, while TEMPO acts as a Lewis base in its reduced state. The polarized water molecules create interactions with the free $H^+$ and $OH^-$ ions that change the chemical activity of these ions in solution. In addition, once free $H^+$ is consumed in the presence of a weak base such as bicarbonates, a hydrogen atom of the polarized water molecule near an oxidized TEMPO provides an $H^+$ and acts as a pH buffer. *In-situ* attenuated total reflectance-Fourier transform infrared spectroscopy (ATR-FTIR) confirms the pH change results in formation and departure of inorganic carbon through redox reactions.

## CO₂ Capture Properties and Mechanism

TEMPO is a stable free radical, stable in both its native (reduced) aminoxyl state as well as its oxidized oxoammonium state (Fig. 2a)[24,25]. The delocalized electron between the N and O that forms a three electron $\pi_{N-O}$ bond stabilizes the molecule, further protected by the steric screening provided by the four adjacent methyl groups[26]. The free radical is readily oxidized and possesses favorable electrochemical properties. TEMPO oxidation is highly reversible, has its redox potential within the thermodynamic stability window of water and has rapid charge-transfer kinetics (Supplementary Fig. 2). We see no TEMPO degradation through 1000 cycles in cyclic voltammetry experiments (Supplementary Fig. 3). In addition, TEMPO-derivatives have molar solubilities in water, making them highly compatible with aqueous solutions[25], and a TEMPO derivative used in a redox flow battery has been shown to have a capacity retention of 99.993% per cycle over 1,000 consecutive cycles[24]. For these reasons, TEMPO-derivatives have been deployed in redox flow batteries[24–29], and as redox mediators in lithium-air batteries[30–32]. In this study, we primarily focus on 4-hydroxy-TEMPO (H-TEMPO) (Fig. 2a), but we also demonstrate the viability of 4-amino-TEMPO (A-TEMPO) (Supplementary Fig. 4-5).

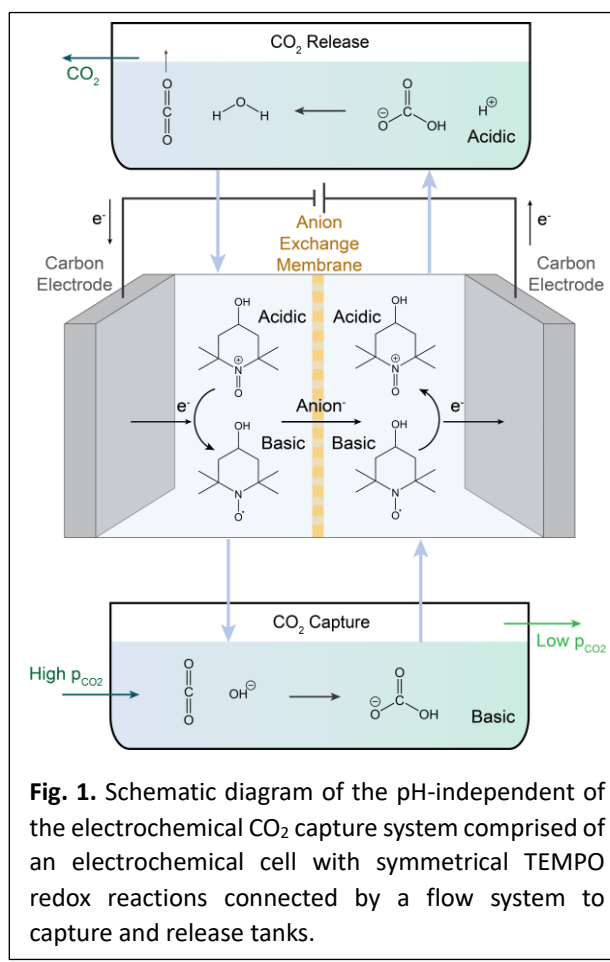

**Fig. 1.** Schematic diagram of the pH-independent of the electrochemical CO₂ capture system comprised of an electrochemical cell with symmetrical TEMPO redox reactions connected by a flow system to capture and release tanks.

Although TEMPO had been used for several applications, to the best of our knowledge, it had not been explored for electrochemical $CO_2$ capture. The $CO_2$ capture capability of TEMPO was tested using an H-cell (Fig. 2b). The electrolyte solution containing 0.4 M of H-TEMPO, 0.2 M KOH and 1.2 M KCl was initially bubbled with 20% $CO_2$ and 80% $N_2$ for 1 hour under open-circuit conditions. Subsequently, the gas inside the cell was flowed at 1 sccm using a mass flow controller and a vacuum pump, and the $CO_2$ concentration of the stream was monitored. At the outset we observe low $CO_2$ concentration in the effluent, which increases to above 90% upon applying an anodic current of 100 mA (Fig. 2c). After the current is stopped, $CO_2$ release is stalled and concentration is dropped. This H-cell experiment demonstrates that the H-TEMPO solution with high pH spontaneously captures $CO_2$ and subsequently releases it upon electrochemical oxidation.

To further investigate the $CO_2$ capture mechanism, we conducted an *in-situ* attenuated total reflectance-Fourier transform infrared spectroscopy (ATR-FTIR) experiment (Fig. 2d). The electrochemical setup is similar to the one depicted in Fig. 2b, with the addition of a $CO_2$ gas micro-bubbler and a peristaltic pump for transport of the TEMPO solution. The TEMPO solution was continuously circulated between the H-cell and a custom designed ATR-FTIR flow cell at 50 mL min$^{-1}$ [33]. Within the flow cell, the TEMPO solution flowed over a Ge ATR crystal, allowing for continuous infrared measurements of the TEMPO solution at the Ge-liquid interface. The experiment begins in its native state, and then undergoes CO$_2$ injection. A peak around 1360 cm$^{-1}$ that corresponds to the bicarbonate anion[34] emerges with CO$_2$ injection, signaling bicarbonate formation (Fig. 2e and Supplementary Figs. 6-8 for A-TEMPO). After reaching a steady state, an oxidative current is turned on while continuing to bubble CO$_2$. The bicarbonate peak decreases with time, consistent with the conversion of bicarbonate into CO$_2$ through a decrease in measured pH of the electrolyte solution. This hypothesis is further substantiated through ultraviolet-visible (UV-Vis) absorption spectroscopy (Supplementary Figs. 9-10) and liquid chromatography-mass spectrometry (LC-MS) (Supplementary Figs. 11-16), which rule out an alternative mechanism of direct absorption of CO$_2$ onto TEMPO.

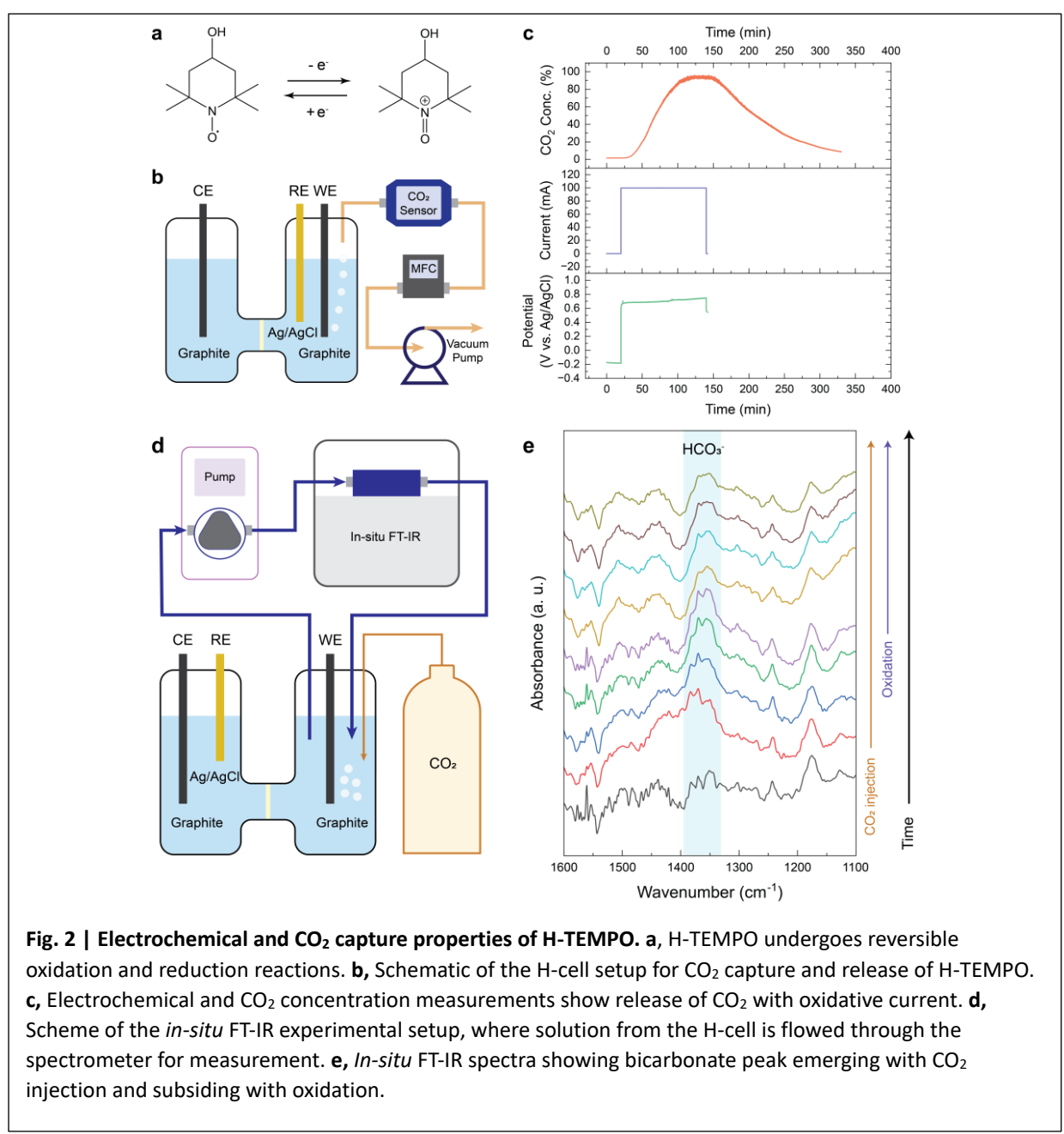

**Fig. 2 | Electrochemical and CO$_2$ capture properties of H-TEMPO. a,** H-TEMPO undergoes reversible oxidation and reduction reactions. **b,** Schematic of the H-cell setup for CO$_2$ capture and release of H-TEMPO. **c,** Electrochemical and CO$_2$ concentration measurements show release of CO$_2$ with oxidative current. **d,** Scheme of the *in-situ* FT-IR experimental setup, where solution from the H-cell is flowed through the spectrometer for measurement. **e,** *In-situ* FT-IR spectra showing bicarbonate peak emerging with CO$_2$ injection and subsiding with oxidation.

**pH modulation mechanism of TEMPO redox**

To test the pH swing hypothesis, we measured the pH of the solution after varying degrees of oxidation of H-TEMPO at 0.4 M concentration (Fig. 3a). We observe that the pH at the reduced state is about 9.4 but drops with oxidation and reaches 1.8 at complete oxidation (Supplementary Fig. 17), and similar pH changes could also be obtained with A-TEMPO (Supplementary Fig. 18). The $pH$ approximately follows the change in $-\log_{10}[TEMPO^+]$, where $TEMPO^+$ (Lewis acid) is the oxidized state of TEMPO, signifying that the pH change scales with the amount of $TEMPO^+$ produced. However, the $[TEMPO^+]$ is ~10-fold higher than the hydrogen activity $\{H^+\} \equiv \gamma[H^+]$, signifying that $TEMPO^+$ acts as a weaker acid than free $H^+$ (Fig. 3a). Although protons or hydroxides do not appear to be directly consumed or produced through TEMPO redox, we observe that the pH changes with TEMPO redox reaction.

Importantly, the thermodynamics of TEMPO redox is independent of pH (Supplementary Figs. 19-20). Fig. 3b shows that the equilibrium potential, which is the midpoint between the peaks of cyclic voltammograms in the inset, remains essentially constant with pH. A similar pH-independent nature can also be observed in A-TEMPO (Supplementary Figs. 21-22) and has also been previously observed in flow battery applications[25,35]. pH-independence is a key property for $CO_2$ capture via electrochemical pH modulation.

For redox reactions that directly modulate $[H^+]$, their thermodynamics change with pH, as it is a direct measure of the chemical potential of $H^+$. Therefore, a high pH for $CO_2$ capture leads to a low redox potential and while a low pH for $CO_2$ release leads to a high redox potential, inducing a thermodynamic energy penalty of 0.059 V per pH. For our method, however, because TEMPO redox can swing the pH without directly producing or consuming $H^+$, its equilibrium potential is independent of pH. In Supplementary Note 2, we argue *that if the voltage needed to oxidized $TEMPO_{red} \rightleftharpoons TEMPO^+ + e^-$ is pH independent, then the pH-related energy penalty of the full roundtrip cycle of $TEMPO_{red} \rightarrow TEMPO^+ \rightarrow TEMPO_{red}$ is zero* (Supplementary Note 2).

To further understand the molecular-level mechanism behind pH-independent modulation of pH, we conducted molecular dynamics (MD) simulations (Supplementary Figs. 23-24). We observe that water molecules dominate the first hydration layer of both reduced and oxidized TEMPO (Supplementary Figs. 25-28). Although $K^+$ is attracted to reduced H-TEMPO and $Cl^-$ is attracted to oxidized H-TEMPO, the large number of water molecules compared to other species leads to water populating the primary hydration shell. Fig. 3c-d show that hydrogen atoms of nearest water molecules interact closely with the oxygen on reduced H-TEMPO, whereas oxygen atoms interact closely with the nitrogen of oxidized H-TEMPO (Supplementary Figs. 29-30).

Density functional theory (DFT) simulations of H-TEMPO in the reduced and oxidized states show that the partial charges on the nitroxyl oxygen and nitrogen atoms shift positively with oxidation (Fig. 3g). Fig. 3e-g and Supplementary Fig. 31 show that the oxidized H-TEMPO is positively charged, with a considerable portion of the charge centered on the nitrogen atom. Fig. 3h shows that the radial distance of water oxygen to nitroxyl oxygen depends on the oxidation state of H-TEMPO.

In our proposed model, the activity coefficient in $\{H^+\} = \gamma_{H^+}[H^+]$ is changed by the oxidation of TEMPO. The positive charge associated with the nitrogen atom polarizes the nearby water molecules so those molecules are slightly more likely to dissociate into $H^+$ and $OH^-$ than unpolarized water molecules. Thus, the nearby water molecules can serve as $H^+$ buffers that enable bicarbonate conversion into carbonic acid (Supplementary Note 3). If $H^+$ is consumed in the reaction $H^+ + HCO_3^- \rightarrow H_2CO_3 \rightarrow CO_{2,\ dissolved} + H_2O$, then a proton on a nearby polarized water molecule dissociates to replace the consumed $H^+$, as summarized by the reaction $TEMPO^+ + H_2O \rightleftharpoons TEMPO^+ \cdot OH^- + H^+$.

Note that oxidized TEMPO modulates the pH by polarizing water molecules that replace the consumed protons, but does not increase the existing [H$^+$]. The buffering capability begins to vanish after the $(10\ mL)(0.4\ M) = 0.004\ moles\ TEMPO^+$ have an $OH^-$ electrostatically bound to each molecule (Fig. 3i). Titration of reduced TEMPO with a strong acid shows that reduced H-TEMPO does not show any buffering activity (Supplementary Fig. 32)

In Supplementary Fig. 33, the FT-IR spectra of various TEMPO solutions are shown. We were searching for an unambiguous signature of an increase of the OH stretch vibration due to the creation of $TEMPO^+ \cdot OH^-$, but the vibration frequency appears to be hidden in the vibrational spectra of the much more abundant water molecules. However, a broad peak around 2900 cm$^{-1}$ emerges for solutions with TEMPO$^+$. We hypothesize that this new peak may indicate a change in the vibrational spectra due to weak interactions between TEMPO$^+$ and H$_2$O since the size of the peak scales with [TEMPO$^+$].

The results of our DFT simulations are consistent with our model that states that oxidized TEMPO acts as a Lewis acid (electron acceptor), which polarizes water molecules to alter the activity of protons[36]. Because [TEMPO] ≫ [H$^+$], subtle changes in the interactions between TEMPO and the aqueous electrolyte have a profound influence on the pH. Since TEMPO redox reaction does not directly involve H$^+$, the thermodynamics of the reaction is independent of pH.

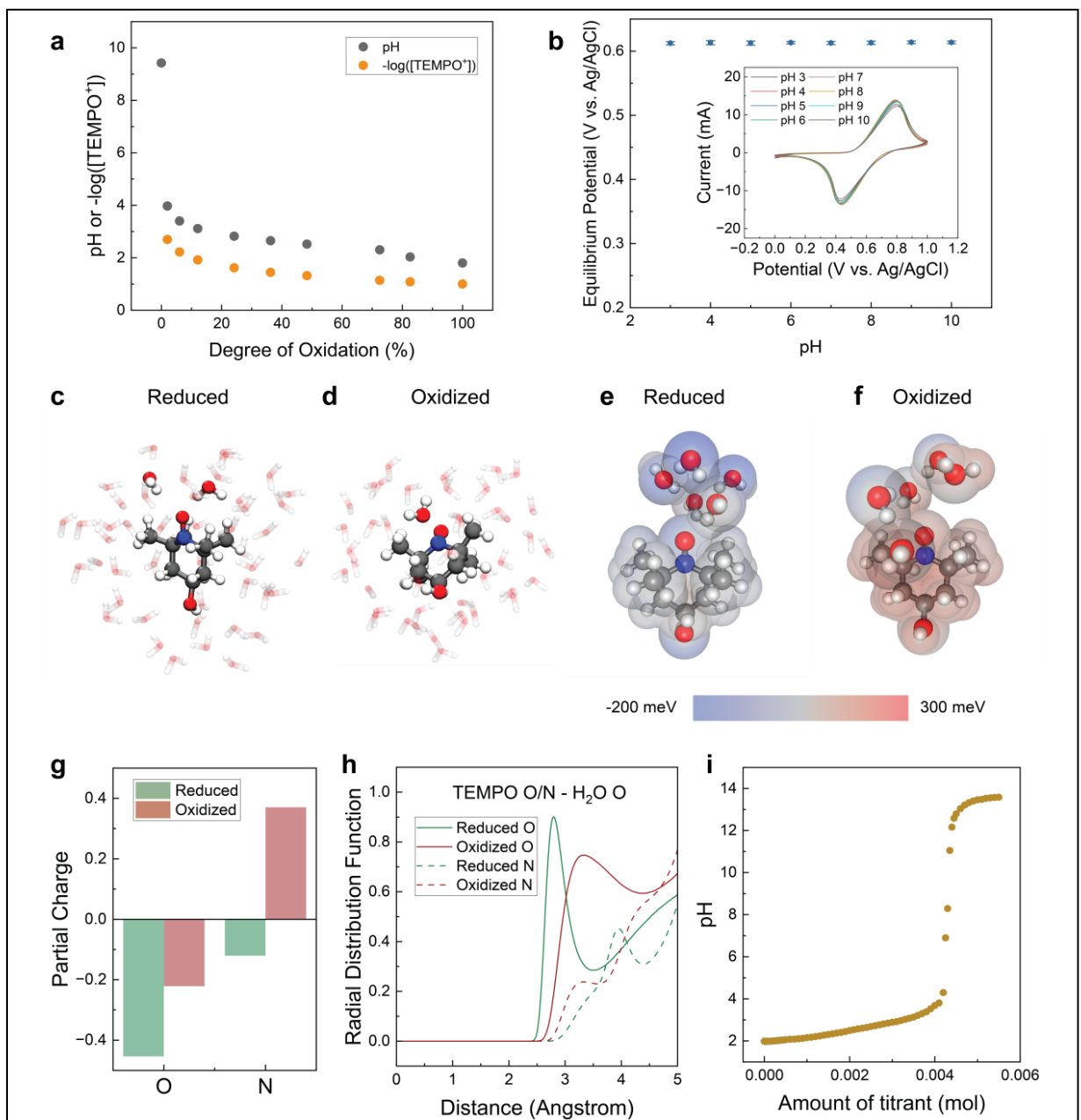

**Fig. 3 | The pH modulation mechanism and pH-independence. a,** The pH of the H-TEMPO solution decreases with oxidation. Note that the molar concentration of $TEMPO^+$ is significantly higher than $H^+$ concentration. **b,** The redox potential of H-TEMPO is largely independent of pH, measured at [TEMPO] = 5 mM. The peak-to-peak potential difference arises from ohmic resistances present in the measurement. **c-d,** MD simulation snapshot of hydration structures of reduced and oxidized H-TEMPO, showing different water polarization. **e-f,** DFT simulation of electrostatic potentials shows that H-TEMPO oxidation positively polarizes 2 water molecules nearest the nitrogen atom. **g,** DFT simulation of the partial charges on nitroxyl N and O on reduced and oxidized H-TEMPO, showing more positive polarization for oxidized TEMPO, and a very slight weakening of the binding energy of the hydrogen atoms to the oxygen atom. **h,** MD simulation of the radial distribution function, defined as $g(r) \equiv \langle \rho(r) \rangle / \rho_{avg}$ where $\rho$ is the density of the oxygen atom attached to water (abbreviated $H_2O$ O) around the nitroxyl O on H-TEMPO. **i,** The buffering capability of $TEMPO^+$ is demonstrated by titrating 1 M KOH into a solution of 10 mL of 0.4 M oxidized H-TEMPO in 1.2 M KCl solution. The pH initially increases slowly with KOH titration, as TEMPO behaves as a weak acid (electron acceptor). When about 4.3 mM of titrant was added, the pH increases rapidly as $TEMPO^+$ is converted into $TEMPO^+ \cdot OH^-$. The pH plateaus at a pH close to that of the KOH titrant.

**CO₂ Capture Performance in a Flow Cell**

To examine the practical efficacy of our TEMPO chemistry, we built a flow cell system that allows for a continuous capture and release of $CO_2$ (Fig. 4a, Supplementary Fig. 34). Our electrochemical cell consists of serpentine flow channels patterned into graphite plates and porous graphite felts as the electrodes separated by an anion exchange membrane. Our $CO_2$ capture and release process has four distinct steps. In step 1, simulated flue gas (20% $CO_2$ 80% $N_2$) is bubbled into the reduced TEMPO solution in the $CO_2$ capture tank. In step 2, the solution is pumped into the anodic chamber, where TEMPO becomes oxidized. In step 3, $CO_2$ is released in the $CO_2$ release tank, where an inert gas carries the gas into the $CO_2$ sensor for concentration quantification. In step 4, the solution after $CO_2$ release is pumped into the cathodic chamber where the TEMPO molecules become reduced and regenerated for a subsequent cycle.

This system configuration has some key advantages. (1) Since the electrochemical cell conducts symmetrical redox reactions, it minimizes the equilibrium cell voltage and thus the energy input. (2) We utilize both half cells to conduct critical steps (i.e. TEMPO oxidation and regeneration) instead of conducting charge-balancing reactions, which reduces materials and capital costs. The anion exchange membrane allows $Cl^-$ transport to maintain charge balance. (3) The continuous capture and release of $CO_2$ maximizes utilization rate and allows for a compact electrochemical $CO_2$ capture system.

Fig. 4b shows the $CO_2$ concentration, current and cell voltage profiles measured for various current set points (See also Supplementary Fig. 35). The experiment is conducted with a solution consisting of 50% reduced TEMPO and 50% oxidized TEMPO. The initial pH of the solution with reduced TEMPO is adjusted by adding 0.2 M KOH. (See Methods, "$CO_2$ capture and release with flow cell" in the Supplementary Information.) $CO_2$ is then flowed for capture prior to applying electrical current. In Fig. 4b and 4c, as the current in full cell goes to zero, the cell voltage difference goes 0 V. With applied current, the cell voltage and $CO_2$ concentration of the release effluent increase. At a relatively low current density of 1.25 mA cm$^{-2}$, the *operando* cell voltage is 0.022 V, which is one of the lowest reported cell voltages for electrochemical $CO_2$ capture[4,5,7,9,16–18,37].

In order to estimate the electrochemical energy required to capture and release $CO_2$, we need to calculate the Faradaic efficiency (FE) of process. We define a FE of 100% as each electron results the capture and release of one $CO_2$ molecule. In Supplementary Note 4, the FE at 1.25 mA cm$^{-2}$ is calculated to be 81.4%, so that the energy cost corresponding to 0.022V is 2.6 kJ mol$^{-1}$ $CO_2$ for the oxidation/reduction roundtrip cycle. Whereas typical pH swing mechanisms require thermodynamic potentials of 0.18-0.65 V that correspond to pH differences of 3-11[5,6,9], our pH-independent redox reaction minimizes equilibrium cell voltages and energy input.

We note here that the low output gas concentrations given in Fig. 4b is low (< 25000 ppm) due to inert gas flow used for $CO_2$ quantification. In a future version of flow cell, we can eliminate the use of an inert carrier gas by simply using a larger electrochemical surface area so that the amount of $CO_2$ released in chamber 3 is sufficient to approach a pressure of 1 atmosphere, as demonstrated in Fig. 2c.

As summarized in Fig. 4c, the overpotential increases linearly with current, which signifies that our electrochemical cell has primarily an ohmic impedance response. This insight is also corroborated by electrochemical impedance spectroscopy (Fig. 4d, Supplementary Fig. 36, Supplementary Note 5).

We also confirmed that the power requirement for electrolyte transport is small compared to the electrochemical counterpart. Fig. 4e-f show the pressure drop across one side of the electrochemical cell for different flow rates and the corresponding power requirements, calculated using the Hagen-Poiseuille equation. For 6.25 mA cm$^{-2}$ operation in our cell, electrochemical power requirement is 890 µW cm$^{-2}$ whereas the pump power requirement for both sides of the cell is 6.9 µW cm$^{-2}$, which is 2 orders of magnitude lower. Fig. 4g shows continuous capture and release of $CO_2$ for 30 hours, confirming that the system can be operated at steady state stably without decrease in Faradaic efficiency and increase in overpotential. We also observe a turnover number, defined as the total moles of released $CO_2$ over the amount of H-TEMPO present, of ~ 1.2, confirming that TEMPO is being reused as a capture and release agent.

While this work focuses on understanding how the pH can be altered by modulating the activity factor, we believe that our $CO_2$ capture performance could be significantly improved. The overpotential of our system could be further decreased since our cell exhibits an ohmic behavior and the electron transfer impedance of $TEMPO_{red} \rightleftharpoons TEMPO^+$ is very low (Fig. 4c-d). . We have tested an anion exchange membrane with 2x lower resistivity. With modest engineering changes in the electrochemical cell, as discussed in Supplementary Fig. 34, we are confident that the electrochemical resistance can be lowered by 3 – 5 fold, since previous researchers have reported redox battery resistances that were 5x lower [28,38]. Additionally, since our redox molecule is very inexpensive, a larger active surface area operating at lower current densities (and less resistive energy costs) are possible.

There is also no need to reduce the pH below 5 since the solubility of inorganic carbon in water becomes negligible. Tuning of the electrolyte and the TEMPO molecular structure could also potentially improve the reduction kinetics. To this end, we have identified variants of H-TEMPO that exhibit pH-independent swings that operate stably up to pH ~ 12, increasing the aqueous absorption capacity to ~ 0.4 moles per liter. Efficient design of capture and release tanks to increase gas-liquid interface while minimizing energy required to expose the $CO_2$ gas to the aqueous solution, and the flow of the liquid in the rest of the system schematically outlined Fig. 1. For example, leveraging the existing knowledge on structured packing and tray engineering will improve liquid and gas mass transfer properties[39–41].

It is important to note that solvation of gaseous $CO_2$ and conversion into bicarbonates is exothermic[42]. Although the decrease in entropy during the process offsets part of the enthalpy change, the total free energy change of the process is negative, providing the thermodynamic driving force. We anticipate that high $CO_2$ capture rates will lead to small increases in temperature of the electrolyte solution absorbing $CO_2$. This excess thermal energy can be used to heat up the solution in the release chamber through heat exchangers to facilitate the release of $CO_2$.

In addition, we believe that TEMPO introduced in this work is only one example of pH-independent redox chemistry. Redox-active molecules that are stable in both reduced and oxidized states, do not directly involve $H^+$, exhibit pH-independence and have high water solubility can be suitable candidates for efficient $CO_2$ capture. For example, we are currently investigating variants of viologens that do not react with oxygen. Viologens are also commonly used in redox flow batteries[27,43]. Other potential materials such as ferrocyanide/ferricyanides and multivalent ions that exhibit pH-independence may also be candidates for efficient $CO_2$ capture.

**Conclusion**

In this study, we demonstrated that pH-independent redox chemistries enable energy-efficient capture and release of $CO_2$. Whereas equilibrium redox potentials depend on the pH in conventional pH swing mechanisms, our redox chemistry is pH-independent by circumventing the direct involvement of $H^+$. The pH-independence of the redox reaction has profound implications in the thermodynamic energy cost of capture and release, as potential gradients that arise from pH are minimized. Through MD and DFT simulations, we show that $TEMPO^+$ acts as a weak Lewis acid that accepts electrons from surrounding water molecules. However, we stress that $TEMPO^+$ does appreciably change the $[H^+]$. Its polarization of water molecules in the near vicinity of nitrogen in the nitroxyl group provides a more readily available source of protons to replace than free water molecules consumed in the reaction $H^+ + HCO_3^- \rightarrow H_2CO_3$. The titration of KOH demonstrates that $TEMPO^+$ interactions with water serve as a powerful buffer and dramatically increases the activity coefficient $\gamma$. We note that even a modest change in the charge distribution of the 1-2 water molecule closest to oxidized nitroxyl group of $TEMPO^+$ can change the activity coefficient since $[TEMPO^+] \gg [H^+]$.

We demonstrated $CO_2$ capture and release through TEMPO redox and investigated the mechanism through spectroscopic techniques including *in-situ* ATR-FTIR characterization. Through a flow cell, we showed negligible equilibrium cell voltage and an *operando* cell voltage as low as 0.022 V, which corresponds to a capture cost of 2.6 kJ mol$^{-1}$ $CO_2$.

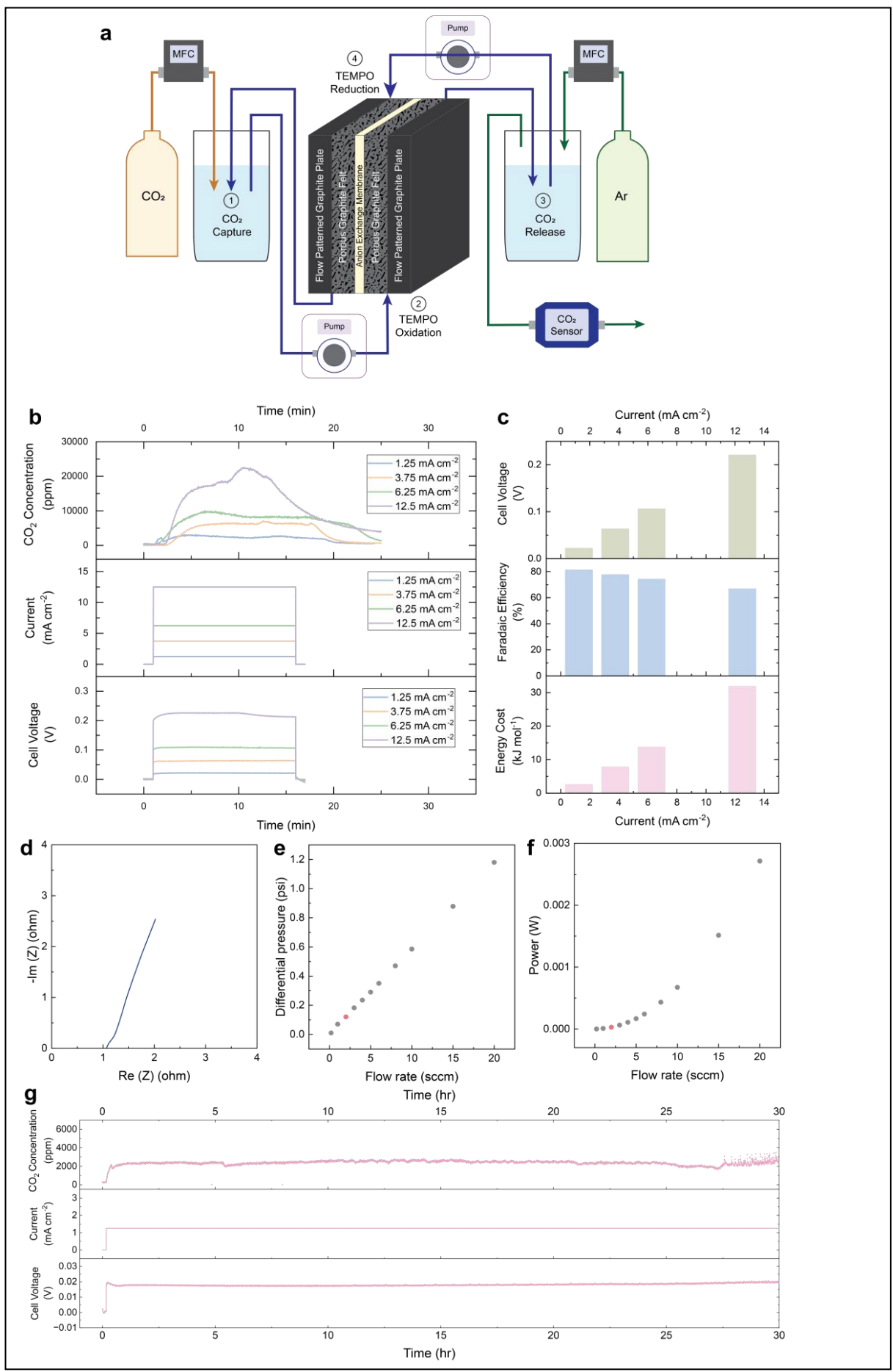

**Fig. 4 | CO₂ capture performance in a flow cell, where [H-TEMPO] = 0.4 M. a,** Schematic diagram of the flow cell system comprised of four distinct steps: 1) $CO_2$ capture; 2) TEMPO oxidation; 3) $CO_2$ release; 4) TEMPO reduction. **b,** $CO_2$ concentration, current and voltage profiles during $CO_2$ capture and release with various currents. **c,** Summary of $CO_2$ capture and release performance including cell voltage, Faradaic efficiency and energy cost. **d,** Electrochemical impedance spectroscopy of the cell. See Supplementary Fig. 36 and Supplementary Note 4 for further details. **e-f,** Differential pressure and pump power across one side of the electrochemical cell with flow rate. **g,** Stability test of the electrochemical $CO_2$ capture system.

**Methods**

Cyclic voltammetry experiments

For all experiments, A-TEMPO (TCI Chemicals) and H-TEMPO (Sigma-Aldrich), potassium chloride (KCl, Sigma-Aldrich) and potassium hydroxide (KOH, Sigma-Aldrich) were used as received. 5 mM of A-TEMPO or H-TEMPO with 100 mM of KCl in deionized water were used for CV experiments. Graphite felt (AvCarb G280A, Fuel Cell Store) was used as working and counter electrodes, while Ag/AgCl reference electrode (Gamry Instruments) was used as the reference electrode in a H-cell (Adams & Chittenden). An anion exchange membrane (FAS-30, Fuel Cell Store) was used between the two half-cells of the H-cell. BioLogic potentiostats were used and a scan rate of 50 mV/s was used. All experiments were conducted at room temperature.

$CO_2$ capture with H-cell

The graphite felt, Ag/AgCl reference electrode, and anion exchange membrane are identical to the CV experiments. An electrolyte solution of 0.4 M H-TEMPO, 1.2 M KCl and 0.2M KOH was used on the working electrode side, while 1.2 M KCl was used on the counter electrode half-cell. Initially, the electrolyte solution was bubbled with 20% $CO_2$ - 80% $N_2$ gas for 1 hour while stirring. A cap sealed with epoxy was used to prevent gas leakage in the half cell while allowing gas flow out of the cell. Gas generated inside the half-cell was flowed out using a mass flow controller (Alicat Scientific) at 1 sccm, whose $CO_2$ concentration is measured using the $CO_2$ sensor (CO2Meter). The pressure in the half cell remained close to 1 atm in this experiment. An electric vacuum pump (Franklin Electric) was used to create the pressure differential for gas flow. After initial resting of 20 minutes, an oxidative current of 100 mA was applied to the working electrode, and $CO_2$ concentration was measured. After 2 hours of oxidation, the current was stopped, while $CO_2$ concentration measurement continued to record the changes in $CO_2$ concentration. All experiments were conducted at room temperature.

LC-MS analysis

Samples were analyzed on two different LC-MS instrument setups: (1) an Agilent 1260 high-performance liquid chromatography (HPLC) instrument paired with an Agilent 6520 accurate-mass quadrupole time-of-flight (Q-TOF) mass spectrometer (6520 LC–MS) or (2) an Agilent 1290 Infinity II UHPLC paired with a coupled Agilent 6546 Q-TOF mass spectrometer (6546 LC–MS). Electrospray ionization (ESI) in positive ionization mode was used for both instruments. For the detection of molecules of interest, HILIC analysis was performed using a Poroshell 120 HILIC-Z column (Agilent, 2.7μm, 2.1 x 100 mm) with water (mobile phase A) and 9:1 ACN:water (mobile phase B), each with 0.1% formic acid and 10 mM ammonium formate, as mobile phases. For 6520 LC-MS, an injection volume of 2 μL with a flow rate of 0.25 mL/min was used with the following 22 min gradient method: 0-3 min, 100% B; 3-13 min 100-60% B; 13-14 min 60-100% B; 14-22 min, 100% B, where, A=(100-B)%. For 6546 LC-MS, an injection volume of 1 μL with a flow rate of 0.25 mL/min was used with the following 12.5 min gradient method: 0-3 min, 100% B; 3-8 min 100-60% B; 8-9 min 60-100% B; 9-12.5 min, 100% B.

For mass spectrometer parameters, the following parameters were used for 6520 LC-MS to collect MS data in positive ion mode: mass range of 50-1000 m/z; drying gas temperature of 300˚C; drying gas flow rate of 10 L/min; nebulizer of 35 psi; fragmentor at 150 V; skimmer at 65 V; Oct 1 RF Vpp at 750 V; VCap at 3500 V; 1000 ms per spectrum). For 6546 LC-MS, samples were run in positive mode (mass range: 30-1700 m/z; drying gas temperature: 325˚C; drying gas flow rate: 10 L/min; nebulizer: 35 psi; sheath gas temperature: 350˚C; sheath gas flow: 12 L/min; fragmentor: 135 V; skimmer: 45 V; Oct 1 RF Vpp: 750 V; VCap: 4000 V; 1000 ms per spectrum. For both instruments, the first 0.5 min of each run was discarded to avoid salt contamination of MS apparatus. For LC-MS data analyses, Agilent MassHunter Qualitative Analysis software was used in general. Extracted ion chromatograms (EICs) shown in figures were generated by extracting for the exact m/z for the target ion of interest with a 20 to 100 ppm mass tolerance.

In-situ FT-IR characterization

ATR-FTIR measurements were taken using a Nicolet IS30 spectrometer and an MCT (mercury cadmium telluride)-A detector at room temperature. The spectrometer was coupled with the VeeMAX III (Pike Technologies) attachment at a 60 degree angle of incidence and 128 scans were added to form each measured spectrum. The custom ATR-FTIR flow cell was constructed to allow liquid flow over a Ge face-angled ATR crystal [33]. A 100 mM H-TEMPO, 300 mM KCl and 80 mM KOH solution was prepared. A Masterflex peristaltic pump (Avantor) was used to pump electrolyte solution at 50 mL min$^{-1}$. First, an initial FT-IR spectrum of the pristine solution was collected, and spectra were collected every 5 minutes. Subsequently, $CO_2$ was bubbled in using a disperser until $CO_2$ saturation. After $CO_2$ saturation, an oxidation current of 3 mA was flowed while the $CO_2$ bubbling continued. Throughout the duration of the experiment, the TEMPO solution was circulated between the ATR-FTIR flow cell and H-cell at 50 mL min$^{-1}$.

Flow cell assembly

The electrochemical flow cell set-up consists of custom machined parts. Stainless steel end plates were used to contain polypropylene frames that served as insulators and hose adaptors to insert the liquid feed tubing. Serpentine flow patterns (1.25 mm for both channel width and depth, 1.25 mm spacing between channels, 8 cm$^2$ of active surface area) were carved on graphite plates (8x8 cm area, 0.25" thickness, McMaster). A protuberance (1x1 cm area) of the graphite plate was used for the connection to the external circuit. Viton gaskets were used to seal the space around 3.1-mm-thick graphite felts (4x4 cm area, Rayon Graphite Felt, Fuel cell store), which were baked in air for 6 h at 400 °C before use. An anion exchange membrane (FAS-30, Fuel Cell Store) was used between the two half-cells of the flow cell. Polyvinylidene fluoride compression fittings were used to connect the cell with Teflon tubing. Two peristaltic pumps (BW100, Chonry) were used to circulate the liquid among the $CO_2$ capture and release tanks. Mass flow controllers (MKS Instruments and Omega Engineering) were used to flow simulated flue gas and argon to the $CO_2$ capture and release tanks, respectively. The gas was dehydrated by passing through a Nafion tubing (CO2Meter) before entering a non-dispersive infrared absorption $CO_2$ sensor (CO2Meter).

$CO_2$ capture and release with flow cell

The solution consisted of 0.4 M H-TEMPO, 1.2 M KCl, and 0.2 M KOH solution. To create a 50% oxidized and 50% reduced mixture, half of the total solution amount was completely oxidized through a constant oxidation current and combined with the unoxidized solution, which was bubbled with $CO_2$ for ≥ 1 hour. The oxidized and unoxidized solutions were mixed and stabilized until equilibrium is reached for 30 minutes. Equal amounts of the solution were allocated to capture and release tanks. 50 mL Falcon tubes (Fisher Scientific) were used as tanks and they were stirred using magnetic stirrers to promote mixing and gas exchange during the experiment. The tanks were sealed properly to prevent gas leakages or mixtures with the atmosphere. An argon carrier gas was flowed at 50 sccm into the release tank, which carried the gas into the $CO_2$ sensor for quantification. Once initial conditions are stabilized, the liquid solution was flowed at 2 sccm and the current was flowed at different current rates for $CO_2$ release. After release, current and liquid solution flow were stopped, while continuing to flow and measure gases. The $CO_2$ concentration was monitored for 9 minutes after the current was stopped to more accurately quantify the released $CO_2$, as $CO_2$ release through bubble formation is not instantaneous and flushing the release tank needs time. The experiments were conducted at room temperature. Electrochemical impedance spectroscopy (EIS) was conducted using a Biologic potentiostat, using a potentiostatic input with frequency ranging from 0.2 MHz to 1 Hz.

Molecular Simulations

Classical, fixed-charge MD was conducted using LAMMPS from initial amorphous configurations. For the two systems that were simulated, boxes containing 4450 water molecules, 32 reduced/oxidized H-TEMPO, 16 $OH^-$, 109 $K^+$, and either 125 or 93 $Cl^-$ anions for the oxidized and reduced H-TEMPO states, respectively. The water molecules were described using the TIP3P forcefield [44], $K^+$, $Cl^-$ and both oxidation states of H-TEMPO were described using the General Amber forcefield with partial charges generated with the AM1-BCC method in ANTECHAMBER [45]. $OH^-$ ions were described using the madrid forcefield [46]. Non-bonded interactions not explicitly specified in the forcefields were generated using Lorentz-Berthelot mixing rules. In all cases, the charges of the ionic species were scaled to 0.75. A 10 Å cutoff for Van der Waals and real space coulomb were applied. Long-range Coulomb interactions were computed with a particle-particle-particle-mesh solver, with an error tolerance of $10^{-3}$. Periodic boundary conditions were also applied in all directions.

For each system, an initial energy minimization at 0 K was performed to obtain the ground-state structure with energy and force tolerances of $10^{-4}$. Then, the system was slowly heated from to 298 K at constant volume over 0.01 ns using a Langevin thermostat, with a damping parameter of 100 ps. 5 cycles of quench-annealing dynamics was then applied in an attempt to eliminate any meta-stable solvation states, where the system temperature was cycled between 298 K and 894 K with a ramp period 0.025 ns followed by 0.1 ns of dynamics at either temperature extreme at constant volume. Next, a constant temperature, constant pressure ensemble was applied using the Andersen barostat for 1.5 ns with a pressure relaxation constant of 1 ps to allow for equilibration of the system density. Finally, we performed 10 ns of constant volume, constant temperature (NVT) production dynamics at 298 K. Radial distribution functions were calculated during this period. Snapshots of the H-TEMPO solvation shells were obtained using Visual Molecular Dynamics (VMD) software using trajectories generated from the production dynamics.

Quantum chemistry simulations were performed using the Q-Chem 5.1 quantum chemistry package on cells comprised of 1 H-TEMPO molecule with 5 waters of random dipole orientation placed around the redox-active nitroxyl moiety. Calculations were conducted at the B3LYP//6-31+G(d,p) level of theory for geometry optimization and the B3LYP//6-311++G** level of theory for single-point energy calculations. No implicit solvent field is applied to either system. Partial charges for each atom were calculated using the CHELPG method native to Q-CHEM [47]. Electrostatic potential maps were generated using the IQ-mol software, also based on the CHELPG method.

**Acknowledgements**

S. C. K. acknowledges support from the Stanford Energy Postdoctoral Fellowship. T.F.J, J.E.M, and J.C. were supported by the U.S. Department of Energy, Office of Science, Office of Basic Energy Sciences, Chemical Sciences, Geosciences, and Biosciences Division, Catalysis Science Program to the SUNCAT Center for Interface Science and Catalysis for *in-situ* FT-IR characterization and analysis.

**Author contributions**

S. C. K. and M. G. contributed equally. S. C. K. and S. C. conceived and designed the investigation. S. C. K. conducted electrochemical property characterization and performance testing. S. C. K. and M. G. designed and fabricated the flow cell. J. H. conducted the MD/DFT simulations. S. C. K., M. G., J. E. M., and J. C. conducted the *in-situ* FT-IR characterization. S. C. K. and Y. D. conducted LC-MS characterization. M. G. conducted the flow pressure drop experiments. T. F. J., Y. C., A. M., Y.-K. T., and S. C. supervised the project. S. C. K. and S.C. co-wrote the paper. All authors discussed the results and commented on the manuscript.

**Competing interests**

A provisional patent application with the findings reported in this work was filed by Stanford University.

**Materials and Correspondence**

Correspondence and requests for materials should be addressed to Steven Chu.

# Supplementary Information

**Electrochemical CO$_2$ capture with pH-independent redox chemistry**

Sang Cheol Kim[1,‡], Marco Gigantino[2,‡], John Holoubek[3], Jesse E. Matthews[2,4], Junjie Chen[2,4], Yaereen Dho[5], Thomas F. Jaramillo[2,4,6], Yi Cui[3,6,7], Arun Majumdar[6,8], Yan-Kai Tzeng[9], Steven Chu[1,6,10*]

[1]Department of Physics, Stanford University, Stanford, California 94305, USA.

[2]Department of Chemical Engineering, Stanford University, Stanford, California 94305, USA.

[3]Department of Materials Science and Engineering, Stanford University, Stanford, California 94305, USA.

[4]SUNCAT Center for Interface Science and Catalysis, SLAC National Accelerator Laboratory, 2575 Sand Hill Road, Menlo Park, CA 94025, USA.

[5]Department of Chemistry, Stanford University, Stanford, California 94305, USA.

[6]Department of Energy Science and Engineering, Stanford University, Stanford, California 94305, USA.

[7]Stanford Institute for Materials and Energy Sciences, SLAC National Accelerator Laboratory, 2575 Sand Hill Road, Menlo Park, California 94025, USA.

[8]Department of Mechanical Engineering, Stanford University, Stanford, California 94305, USA.

[9]SLAC-Stanford Battery Center, SLAC National Accelerator Laboratory, 2575 Sand Hill Road, Menlo Park, California 94025, USA.

[10]Department of Molecular and Cellular Physiology, Stanford University, Stanford, CA 94305, USA.

‡ These authors contributed equally

Correspondence to: S. C. schu@stanford.edu

**Supplementary Note 1. Minimum free energy of CO$_2$ capture and release**

The free energy of CO$_2$, or any gaseous species, can be defined as $G = G^0 + RT\ln(p)$, where G is the Gibbs free energy, $G^0$ is the Gibbs free energy at standard conditions, R is the gas constant, T is temperature and p is partial pressure. Capturing CO$_2$ at a low concentration and releasing at a higher concentration has a change in free energy given by $\Delta G = RT\ln(p_f) - RT\ln(p_i) = RT\ln\left(\frac{p_f}{p_i}\right)$, where $p_f$ is the final pressure and $p_i$ is the initial pressure. As $G^0$ serves as a reference point and is canceled in the subtraction process, we can assume that $G^0 = 0$ for purposes here. Supplementary Fig. 1 graphically illustrates the free energy dependence on partial pressure.

Let us take 1 atm as the final pressure ($p_f = 1$), which represents a pure stream of CO$_2$ at atmospheric pressure. At this pressure, $G = 0$, simplifying the mathematics. Now, the minimum $\Delta G$ depends on the CO$_2$ partial pressure of the initial state ($p_i$), the incoming flue gas or air. A flue gas of 20% will require a minimum energy of 4.0 kJ/mol while air with 400 ppm of CO$_2$ will require a minimum energy of 19.1 kJ/mol. The minimum energy required will also depend on the depth of capture. For example, taking a 20% flue gas and capturing CO$_2$ until the outgoing gas concentration is 2% will require less average energy than if the outgoing gas is 0.5%. To calculate the average free energy requirement for given incoming and outgoing gas concentrations, one must take the integral of the function and normalize by the range.

$$\Delta G_{avg} = \frac{-RT \int_{p_{i2}}^{p_{i1}} \ln(p_i) dp}{p_{i1} - p_{i2}}, \quad (1)$$

where $p_{i1}$ is the incoming gas concentration and $p_{i2}$ is the outgoing concentration. Solving the integral analytically,

$$\Delta G_{avg} = \frac{-RT[p \cdot \ln(p) - p]_{p_{i2}}^{p_{i1}}}{p_{i1} - p_{i2}}. \quad (2)$$

Supplementary Table 1 below summarizes the minimum energy requirements for a set of reference cases. An incoming flue gas stream can range from ~3% (natural gas power plants) to >30% (cement plants). As representative points, we picked 20%, 10% and 5% as incoming gas concentration. Typically, for point source CO$_2$ capture, 90% to 95% of the CO$_2$ present in the flue gas is captured, results of which are summarized below. It is worth noting that from the environmental perspective, controlling the outgoing gas concentration may be advantageous over controlling the capture ratio.

**Supplementary Table 1.** Minimum free energy requirement for incoming and outgoing CO$_2$ concentrations.

| Incoming gas concentration | Outgoing gas concentration | Minimum energy requirement (kJ/mol) |
|---|---|---|
| 20% | - | 4.0 |
| 400 ppm | - | 19.4 |
| 20% | 2% | 5.8 |
| 20% | 1% | 6.1 |
| 10% | 1% | 7.6 |
| 10% | 0.5% | 7.8 |
| 5% | 0.5% | 9.3 |
| 5% | 0.25% | 9.5 |

**Supplementary Note 2. pH-independence of the TEMPO reaction**

Let us consider an electrochemical reaction $Ox^+ + ze^- \rightleftharpoons Red$, where $z$ is the number of electrons transferred in the reaction. At chemical equilibrium, the concentrations are given by the equilibrium constant $K$ of the half reaction $K = a_{red}/a_{ox}$, where $a_{red}, a_{ox}$ are the chemical activities, $a_{ox} = \gamma_{ox}[Ox]$ and $a_{red} = \gamma_{red}[Red]$, and $\gamma_{ox}, \gamma_{red}$ are the activity coefficients.

The Gibbs Free energy is then given by

$$\Delta G = \Delta G^0 + RT \ln(\gamma_{red}[Red]/\gamma_{ox}[Ox]) = \Delta G^0 + RT \ln(a_{red}/a_{ox}), \tag{1}$$

where $R$ is the gas constant, $R = 8.31\ J \cdot mol \cdot K^{-1}$, and $\Delta G^0$ is the free energy change at standard conditions.

The free energy change of the electrochemical reaction ($\Delta G$) is equal to the negative of the electrochemical potential ($E$) multiplied by Faraday constant ($F = 96{,}500\ C \cdot mol^{-1}$) and the number of charges ($z$)

$$\Delta G = -zFE, \tag{2}$$

Thus, the electrochemical potential of the half-cell equals to

$$E_{red} = E_{red}^0 - (RT/zF) \ln a_{red}/a_{ox}, \tag{3}$$

where $E_{red}^0$ is the standard reduction potential. Equation 3 is the Nernst Equation.

For a half cell electrochemical reaction that directly involves $H^+$

$$aA + bB + hH^+ + ze^- \rightleftharpoons cC + dD, \tag{4}$$

the electrochemical potential of the reaction is

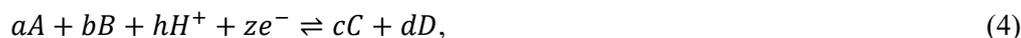

$$E_{red} = E_{red}^0 - (RT/zF) \ln\left(\frac{\{C\}^c\{D\}^d}{\{A\}^a\{B\}^b\{H^+\}^h}\right) = E_{red}^0 - \frac{0.05916}{z}\log\left(\frac{\{C\}^c\{D\}^d}{\{A\}^a\{B\}^b}\right) - 0.05916\left(\frac{h}{z}\right)pH \tag{5}$$

The curly brackets indicate the activities of the reactants, and a change of basis $\ln x = 2.303 \log_{10} x$ was used to obtain $2.303(RT/zF) = 0.05916/z$, where T=298 K. The $H^+$ term was isolated as the third term on the right side of Eq. 5, where $pH = -\log\{H^+\}$.

In the case where $H^+$ is directly involved in the electrochemical reaction, pH changes with the amount of $H^+$ produced or consumed, and there is a -0.0592 V dependence on $pH$. The $pH$ of water is defined as $pH \equiv -\log_{10} a_{H^+} = -\log_{10} \gamma[H^+]$, where the proton chemical activity is $a_{H^+} = \gamma[H^+]$.

In the case of our TEMPO redox reaction $TEMPO^+ + e^- \leftrightarrow TEMPO$, we change the $pH$ not by directly modulating [H$^+$] but rather by changing the activity coefficient $\gamma$. In the cyclic voltammetry data in Fig. 3b, between a $pH$ change of 3 to 10, the center voltage position where the reduced and oxidized states are in equilibrium is ~ 0.62 V. The slope of the $pH$ dependence is $+5.5 \pm 6.38 \times 10^{-5}$ (Supplementary Fig. 20).

Using the stringent bounds on the slope of the $pH$ dependence, we now derive an upper limit on the contribution of $H^+$ that can arise from the reaction

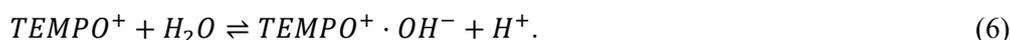

$$TEMPO^+ + H_2O \rightleftharpoons TEMPO^+ \cdot OH^- + H^+. \tag{6}$$

If the TEMPO directly increases $[H^+]$ in Eq. 6, the pH dependence would have a slope of $-0.0592\ V/pH$. However, a pH dependence of $-0.0592\ V/pH$ is 925 σ away from the observed slope of $+5.5 \pm 6.38 \times 10^{-5}\ V/pH$. Therefore, we can rule out the reaction in Eq. 6 based on our pH dependence data in Fig. 3b.

Quantitively, assume that a fraction of $H^+$ is formed in the coupled reaction

$$TEMPO_{red} \rightleftharpoons (1-n)TEMPO^+ + n(TEMPO^+ \cdot OH^-) + n(H^+) + e^-. \qquad (7)$$

If this coupling is correct, suppose that for each $TEMPO^+$, a small number $n < 1$ of $TEMPO^+$ produces a proton $H^+$.

$$\Delta E = -0.05916 \log([TEMPO^+]^{1-n}[TEMPO^+ \cdot OH^-]^n[H^+]^n/[TEMPO_{red}]) \qquad (8)$$

$$\Delta E \sim -\frac{0.05916}{z}(2n)\log[H^+] = -0.118n\log[H^+]. \qquad (9)$$

Note that the slope of the voltage needed to oxidize $TEMPO_{red} \rightarrow TEMPO^+ + e^-$ is $(5.5 \pm 6.38) \times 10^{-5}$ per $\Delta pH = 1$ and per mole of TEMPO, whereas the slope should have been negative. The **1σ uncertainty** of having the needed *negative* slope is $(5.5 - 6.38) \times 10^{-5} = -8.8 \times 10^{-6}$ per $\Delta pH = 1$. The **3σ uncertainty** is $-1.36 \times 10^{-4}$ per $\Delta pH = 1$.

Solving for $n$ in Equation 9, $-0.118n = -8.8 \times 10^{-6}$, $n = 7.46 \times 10^{-5}$ mole $H^+$ per mole of TEMPO per unit change of *pH*. This is a small change in $H^+$ compared to the observed change in the $\Delta pH = 9.4 \rightarrow 1.8 = 7.6$ when fully reduced TEMPO is converted to fully oxidized TEMPO. Thus, the pH change of $9.4 \rightarrow 1.8$ cannot be accounted for by the direct $H^+$ formation due to the reaction in Equation 6.

We are led to conclude that the number of hydrogen atoms directly released from water molecules polarized by TEMPO is negligible compared to the change in $\Delta pH$, and the entropy penalty of changing $[H^+]$ is also negligible.

**Supplementary Note 3. pH modulation mechanism of TEMPO**

In Supplementary Note 2 we provided details on how TEMPO redox does not directly modulate $[H^+]$. Instead, TEMPO changes the activity coefficient of $H^+$ by further polarizing the surrounding water molecules. As shown in Fig. 3c-3f, water molecules around reduced TEMPO are negatively polarized while those around oxidized TEMPO are positively polarized. Hydrogen atoms attached to positively polarized $H_2O$ are pointing away from $TEMPO^+$, and are *very slightly easier to detach* in the reaction $TEMPO^+ + H_2O \rightarrow TEMPO^+ \cdot OH^- + H^+$, where $TEMPO^+$ and $OH^-$ are weakly interacting.

In the presence of weak bases such as bicarbonates, existing $H^+$ will be consumed in the reaction to form carbonic acid. In our model, the more easily detachable hydrogen atoms serve as a buffer source of protons, but they remain attached to water until a $H^+$ is used in the reaction $HCO_3^- + H^+ \rightarrow H_2CO_3$. These polarized water molecules act as a buffer that maintains the proton concentration and the conversion of bicarbonate to neutral carbonic acid and then eventually to $CO_{2,gas}$ in the reactions

$$\begin{aligned}TEMPO^+ + H_2O + HCO_3^- + H^+ &\rightleftharpoons TEMPO^+ \cdot OH^- + H^+ + H_2CO_3 \\ &\rightleftharpoons TEMPO^+ \cdot OH^- + H^+ + CO_{2,aq} + H_2O \\ &\rightleftharpoons TEMPO^+ \cdot OH^- + H^+ + CO_{2,gas} + H_2O. \end{aligned} \qquad (10)$$

This is due to Le Chatelier's Principle, which tells us that a net reaction will occur in the direction that will partially counteract this change and return to the stable equilibrium. $[H^+]$ will be restored from the positively polarized $H_2O$, which splits into $H^+$ and $OH^-$ that is weakly interacting with $TEMPO^+$. We believe that TEMPO-induced water polarization, which modulates the activity coefficient of $H^+$ and enables the buffering of $H^+$, is the mechanism of how TEMPO modulates the pH in our system.

**Supplementary Note 4. Faradaic efficiency and energy cost calculations**

In our flow cell, Faradaic efficiency is calculated by taking the ratio between moles of $CO_2$ released and moles of electrons flowed in the electrochemical cell. Here we will take the 1.25 mA cm$^{-2}$ case shown in Fig. 4b-c as an example to illustrate the calculations. First, we take the total amount of $CO_2$ released by using the following expression: $\int_{t_i}^{t_f}(C - C_i)\rho r \, dt$. Here, $t_i$ is initial time (t=0 min), $t_f$ is final time (t=25 min), C is $CO_2$ concentration, $C_i$ is initial concentration at t=0 min, $\rho$ is molar density at standard temperature and pressure (1 mol/22.4 liters), r is flow rate (50 cc/min). Here, although current is stopped at t=16 min, we take the final time as t = 25 min and record the $CO_2$ released for another 9 minutes without current, because $CO_2$ release is not instantaneous (i.e. bubble formation and escape takes time) and flushing the tank takes time. For the 1.25 mA cm$^{-2}$ case, we obtain 7.63 x 10$^{-5}$ mole of $CO_2$ released.

The moles of electrons flowed can be calculated by the following expression: $JA\Delta t/F$. Here, F is the Faraday constant (96485 C/mol), J is the current density (1.25 mA cm$^{-2}$ for this example), A is active surface area (8 cm$^2$), and $\Delta t$ is the time of current flow (15 minutes). The calculation yields 9.33 x 10$^{-5}$ moles of electrons. Taking 7.63 x 10$^{-5}$ moles of $CO_2$ divided by 9.33 x 10$^{-5}$ moles of electrons gives a Faradaic efficiency of 81.4%. The energy cost can then be calculated by taking the average cell voltage during current flow (0.022 V) and multiplying it by the Faraday constant (96485 C/mol) and dividing by the Faradaic efficiency (81.4%), which yields 2.6 kJ mol$^{-1}$.

**Supplementary Note 5. Impedance analysis of the electrochemical cell in the flow system**

An electrochemical system can be modeled as an equivalent circuit and electrochemical processes can be modeled as circuit components. Resistive processes, including electron and ion transport, can be modeled as a resistor. Processes at an electrochemical interface can be divided into two parallel processes: a Faradaic process and a non-Faradaic process. The non-Faradaic process involves ion adsorption onto the electrode surface in the electrical double layer and can be modeled as a capacitor. The Faradaic process involves charge-transfer reactions that can be modeled as a resistor and a Warburg impedance component that accounts for the changes in concentrations of reactants and products. The impedances of these components will constitute the impedance of the equivalent circuit, whose real part will be the resistance of the electrochemical cell. Lower resistance will translate to lower overpotential and higher energy efficiency. Additional background information on electrochemical impedance spectroscopy can be found in these articles.[1,2]

The equivalent circuit of the electrochemical cell in our flow system is shown in Supplementary Fig. 36a. $R_{ohm}$ is the ohmic resistance of electron and ion transport, $R_{ct}$ is the charge-transfer resistance, $C_{dl}$ is the capacitance of the double layer, and $Z_W$ is the Warburg impedance. The two interfaces in our electrochemical cell are symmetric and therefore are modeled as one. Such an equivalent circuit will result in a Nyquist plot shown in Supplementary Fig. 36b. At high frequencies, the resistance of the cell is $R_{ohm}$. The semicircle represents the charge-transfer process, and the diameter of the semicircle is $R_{ct}$. At low frequencies, Warburg impedance due to concentration gradients adds to the resistance, represented by the slope on the right side of the semicircle.

Fig. 4d shows the electrochemical impedance spectroscopy (EIS) results of our electrochemical cell. We observe that the semicircle is incomplete and proceeds as a slope that corresponds to the Warburg impedance, possibly because the time constant of the RC circuit $\tau \equiv R_{ct}C_{dl}$ overlaps with the time needed for concentration gradient formation. The Warburg impedance is dependent on convection of the electrolyte and a faster pump speed can reduce the Warburg impedance. Importantly, we observe that $R_{ct}$ is much smaller than $R_{ohm}$, which signifies that the intrinsic reaction kinetics of TEMPO contributes only a small fraction of the total resistance of the cell. $R_{ohm}$ can be reduced through improvements in anion exchange membranes and reductions in electrode thickness and contact resistances. We believe that smaller ohmic resistance can be achieved, as they have already been demonstrated in existing redox flow battery literature.[3,4]

**Supplementary Figures**

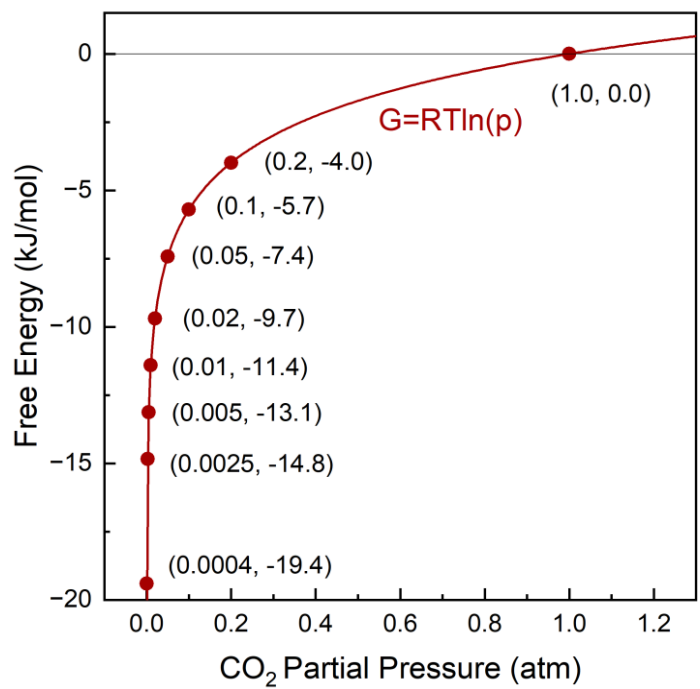

**Supplementary Fig. 1.** Free energy of $CO_2$ dependence on partial pressure. Minimum free energy required for $CO_2$ concentration is $\Delta G = RT \ln\left(\frac{p_f}{p_i}\right)$, which can be calculated by taking the free energy difference between the initial and final partial pressures. For the case where the final pressure is 1 atm, the negative of the free energy at the initial pressure will constitute the minimum free energy required. The integral of the curve normalized by the concentration range yields the average free energy requirement for given incoming and outgoing concentrations. Supplementary Note 1 describes the details of the calculations.

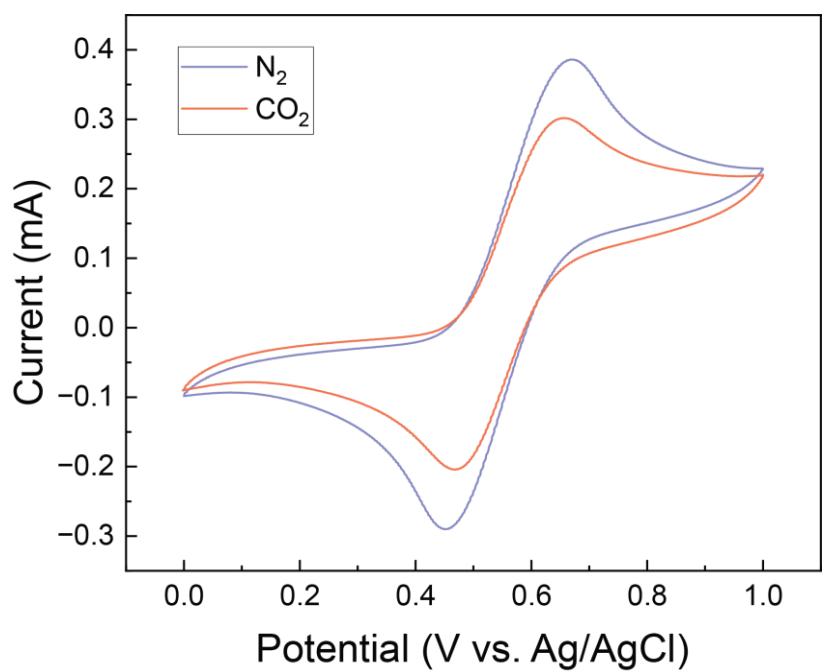

**Supplementary Fig. 2.** Cyclic voltammetry of 5 mM of H-TEMPO with 100 mM of KCl under $N_2$ or $CO_2$ bubbling. Voltage is swept at 50 mV s$^{-1}$ and the peak with the positive current corresponds to the oxidation reaction while the peak with the negative current corresponds to the reduction reaction. We see that in both cases TEMPO redox is reversible and the peak positions are almost identical. This is important because it means that the equilibrium potential does not change when $CO_2$ is injected, which suggests that the thermodynamic energy penalty associated with $CO_2$ capture is minimal. The gap between the oxidation and reduction peaks is pertinent to the particular experimental setup, such as the ohmic drop in the H-cell and small active surface area and is not an intrinsic property of TEMPO. Note that a low concentration of 5 mM was used to ensure that a peak emerges due to mass transport limitations. The initial increase in current signifies an electrochemical oxidation reaction that consumes current, but current drops because of depletion of reactants and a limitation of mass transport. More information on cyclic voltammetry can be found in this introductory paper[5].

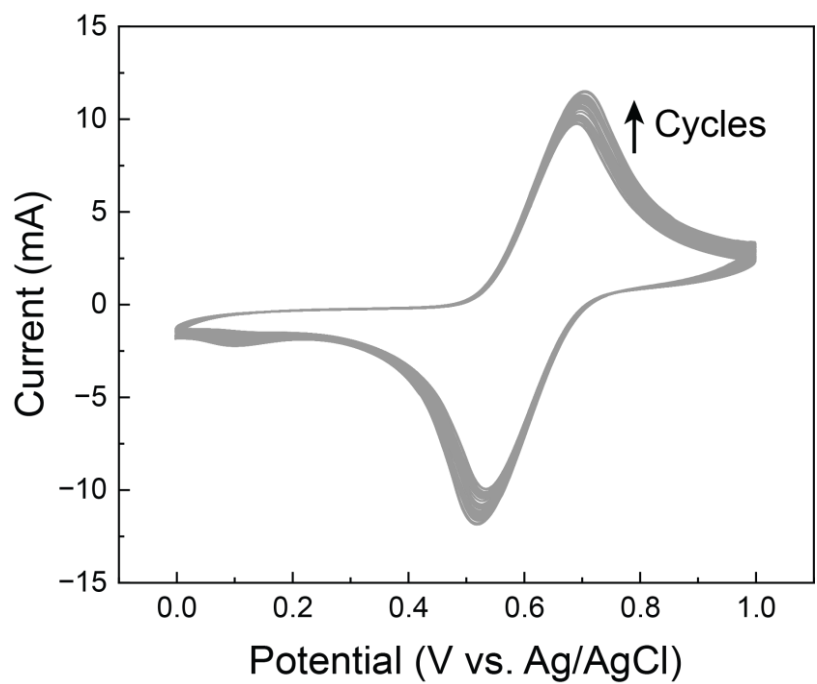

**Supplementary Fig. 3.** Cyclic voltammetry of 5 mM of H-TEMPO with 100 mM of KCl repeated for 1000 cycles at 50 mV s$^{-1}$ sweep rate under $CO_2$ bubbling conditions. The peak currents remain relatively steady over 1000 cycles; in fact, they slightly increase over cycles, potentially due to improved wetting of porous electrodes. This result demonstrates that H-TEMPO redox is highly stable. The gap between the oxidation and reduction peaks is pertinent to the particular experimental setup, such as the ohmic drop in the H-cell and small active surface area and is not an intrinsic property of TEMPO.

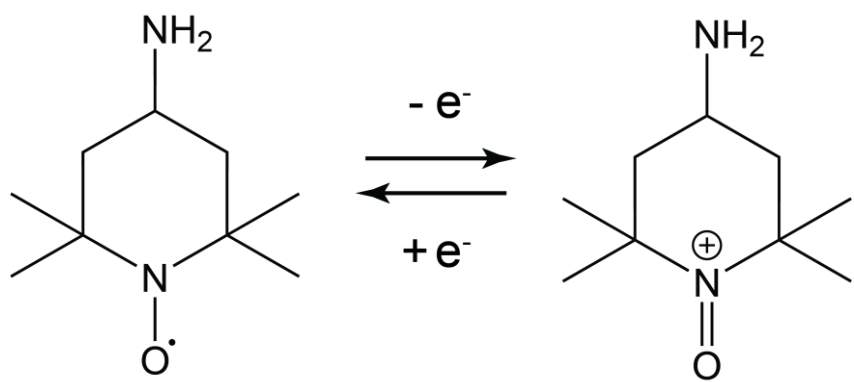

**Supplementary Fig. 4.** Reversible oxidation and reduction reactions of 4-amino-TEMPO (A-TEMPO). It is a stable free radical in its reduced form, stabilized by delocalized electrons between nitroxyl nitrogen and oxygen. Through oxidation, the positively charged oxoammonium is formed, which is also stable in aqueous solutions.

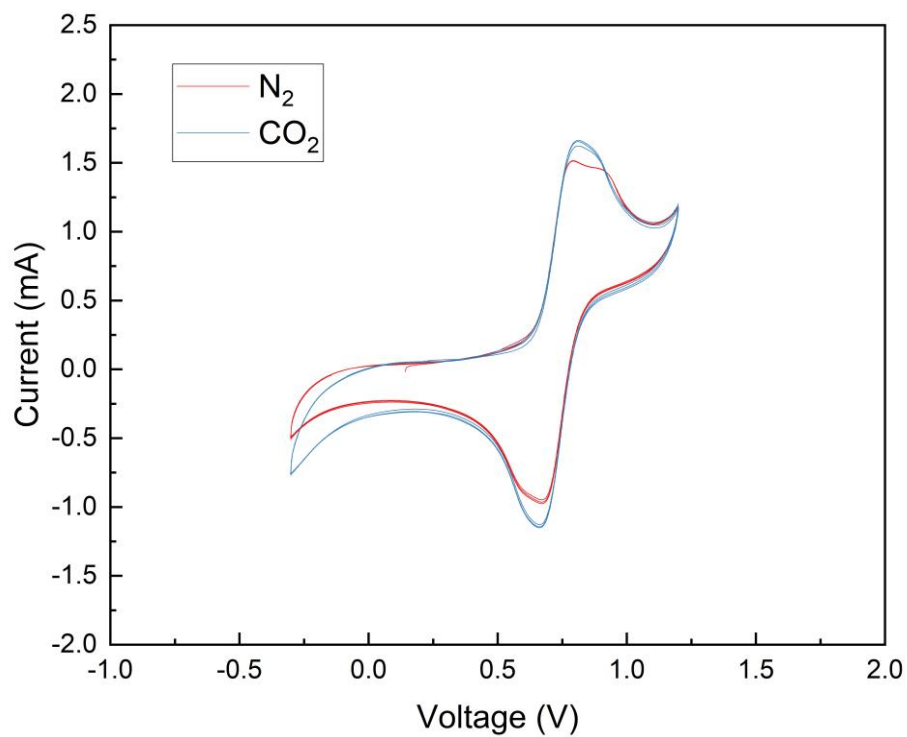

**Supplementary Fig. 5.** Cyclic voltammetry of 5 mM A-TEMPO 100 mM KCl solution under $N_2$ and $CO_2$ bubbling, showing similar peak positions and equilibrium potential. This result is similar to H-TEMPO, suggesting minimal thermodynamic energy difference with and without $CO_2$. The gap between the oxidation and reduction peaks arises due to kinetic factors in the particular H-cell setup and is not an intrinsic property of A-TEMPO.

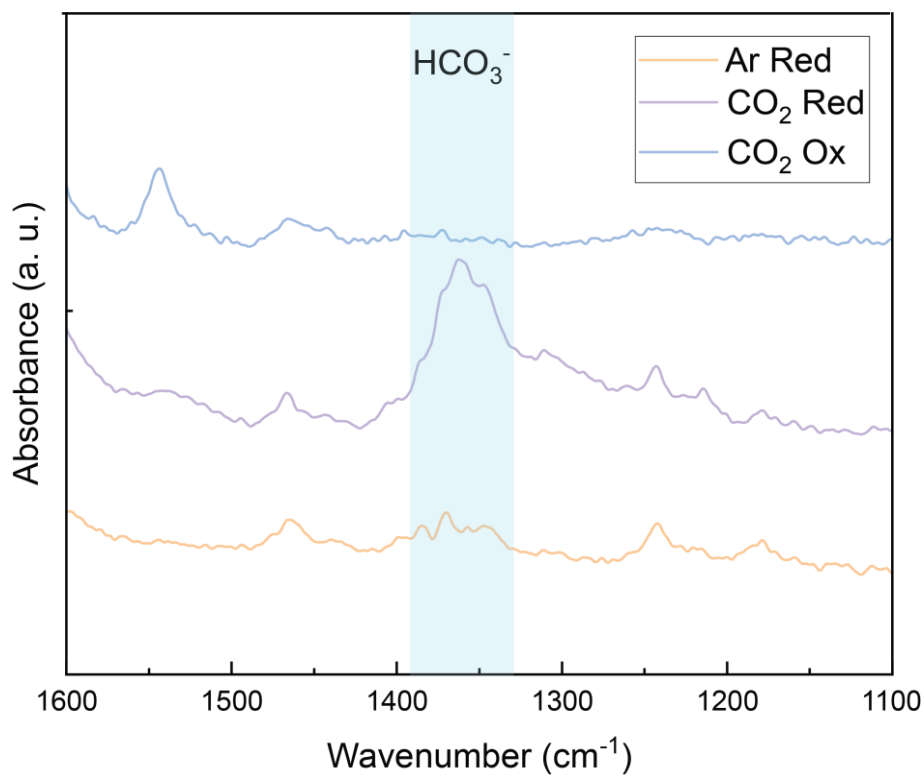

**Supplementary Fig. 6.** Ex-situ FT-IR of 100 mM A-TEMPO 400 mM KCl under Ar, $CO_2$ and oxidized A-TEMPO with $CO_2$, showing that bicarbonate peak around 1360 cm$^{-1}$ appears for $CO_2$ bubbled reduced solution but disappears after oxidation. This shows that A-TEMPO can capture $CO_2$ in the form of bicarbonate, but oxidation of A-TEMPO converts bicarbonate into $CO_2$ that is released. This is because TEMPO oxidation lowers the pH and increases the activity of H$^+$, which combines with bicarbonate to form carbonic acid. Carbonic acid is subsequently dissociated into $H_2O$ and $CO_2$, and the resulting $CO_2$ is released.

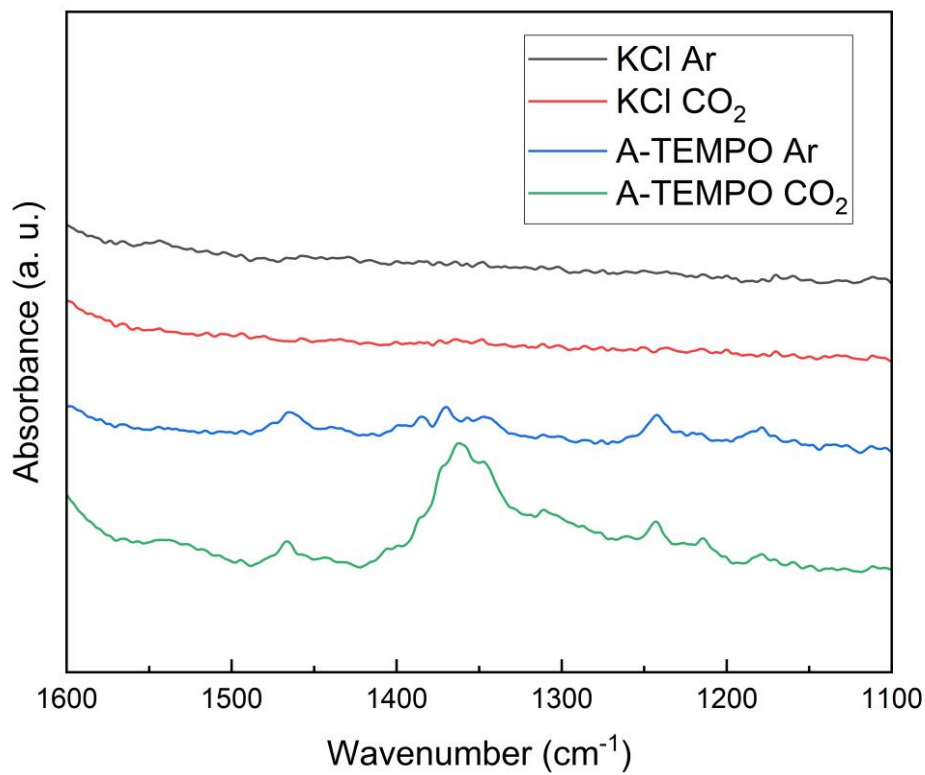

**Supplementary Fig. 7.** *Ex-situ* FT-IR of 400 mM KCl electrolyte without A-TEMPO and 100 mM A-TEMPO 400 mM KCl solution with and without $CO_2$ injection, showing that bicarbonate peak around 1360 cm$^{-1}$ emerges only with TEMPO and $CO_2$ injection. A-TEMPO is a weak base that allows for bicarbonate formation with $CO_2$ injection. However, bicarbonate is not formed without A-TEMPO and $CO_2$ bubbling.

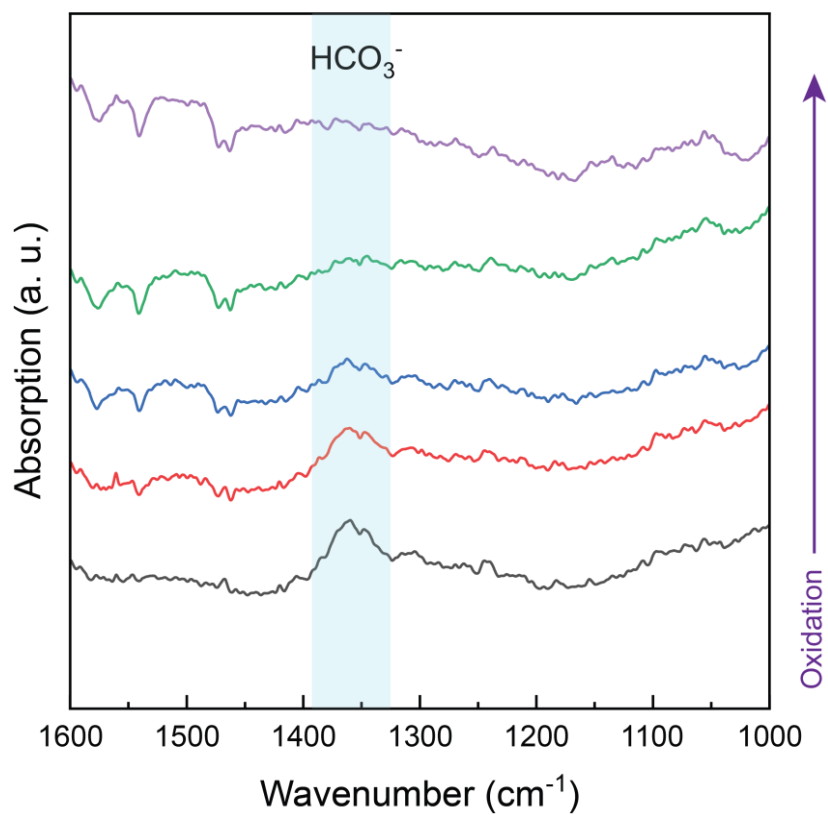

**Supplementary Fig. 8.** *In-situ* FT-IR of 5 mM A-TEMPO 100 mM KCl solution, which starts out saturated with $CO_2$ and is oxidized over time. The bicarbonate peak around 1360 cm$^{-1}$ subsides over time with oxidation. At the onset of the experiment, A-TEMPO is a weak base that allows for the capture of $CO_2$ in the form of bicarbonates. Therefore, a strong bicarbonate peak is shown. However, with oxidation of A-TEMPO, the bicarbonate peak subsides over time. This is because oxidation decreases the pH, which causes the conversion of bicarbonate into carbonic acid, which subsequently dissociates into $H_2O$ and $CO_2$.

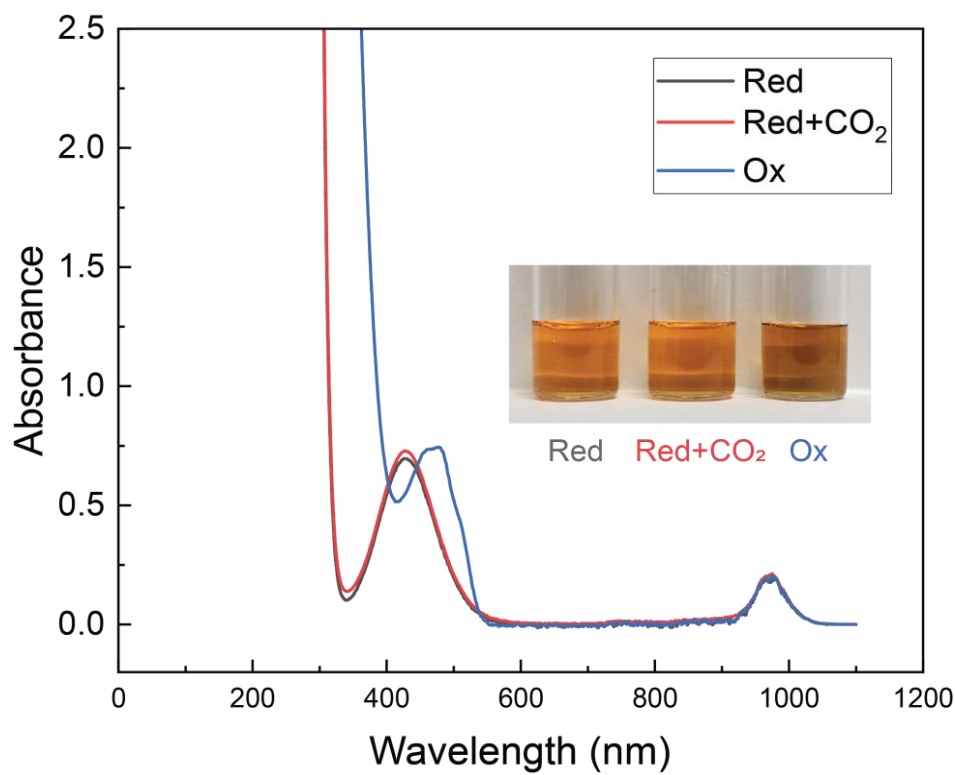

**Supplementary Fig. 9.** Ultraviolet-visible (UV-Vis) absorption spectra of the reduced state, the reduced state with $CO_2$ injected, and the oxidized state of 100 mM H-TEMPO in 400 mM KCl. The reduced sample, both before and after $CO_2$ injection, have absorption peaks around 510 nm, which gives rise to its orange color. The absorption spectrum redshifts after oxidation, which is also reflected in the color change. However, absorption properties remain identical with $CO_2$ injection, which hints that the molecular structure of TEMPO likely does not change. This data is consistent with our proposed $CO_2$ capture mechanism that $CO_2$ does not directly absorb to TEMPO but is captured through bicarbonate formation.

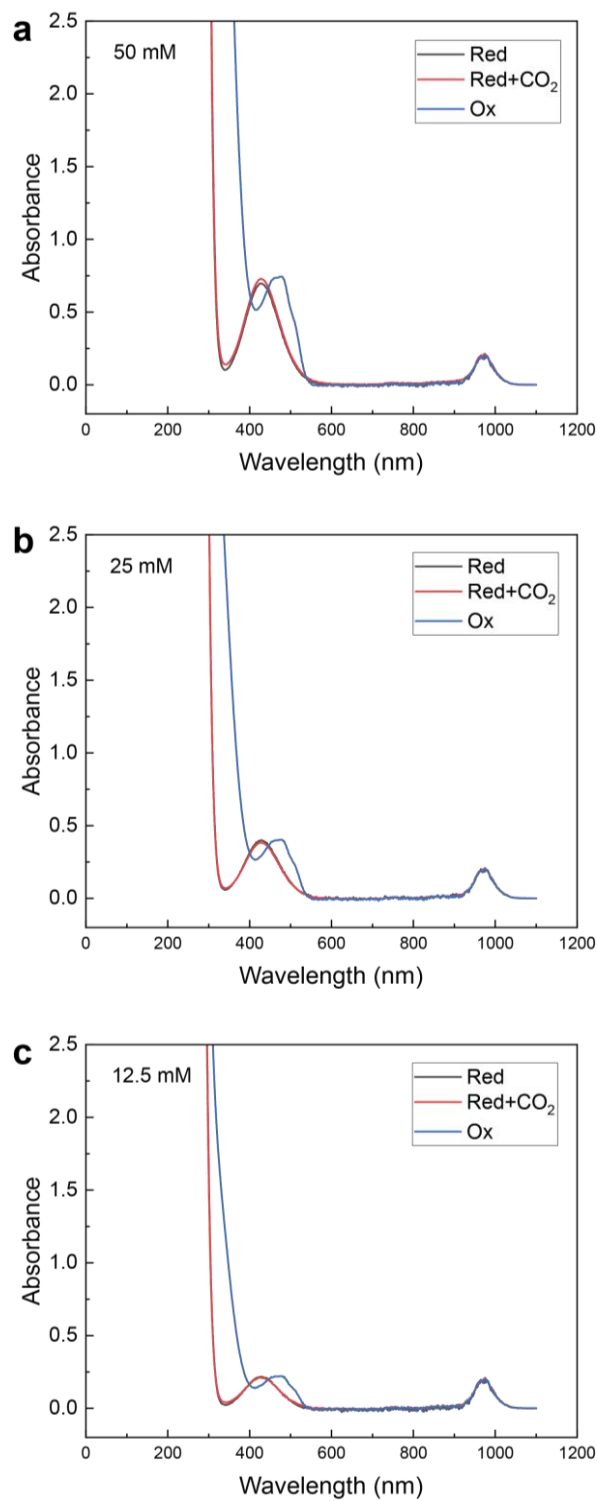

**Supplementary Fig. 10.** UV-Vis spectra of reduced, $CO_2$ bubbled in when TEMPO is in the reduced state, and oxidized H-TEMPO at different concentrations: **a,** 50 mM; **b,** 25 mM; **c,** 12.5 mM. With decreased concentration absorption intensities decrease but the peak positions remain the same. Absorption spectra from various concentration also suggest that the molecular structure of reduced TEMPO likely does not change after $CO_2$ injection. This result provides support that $CO_2$ does not directly bind to TEMPO.

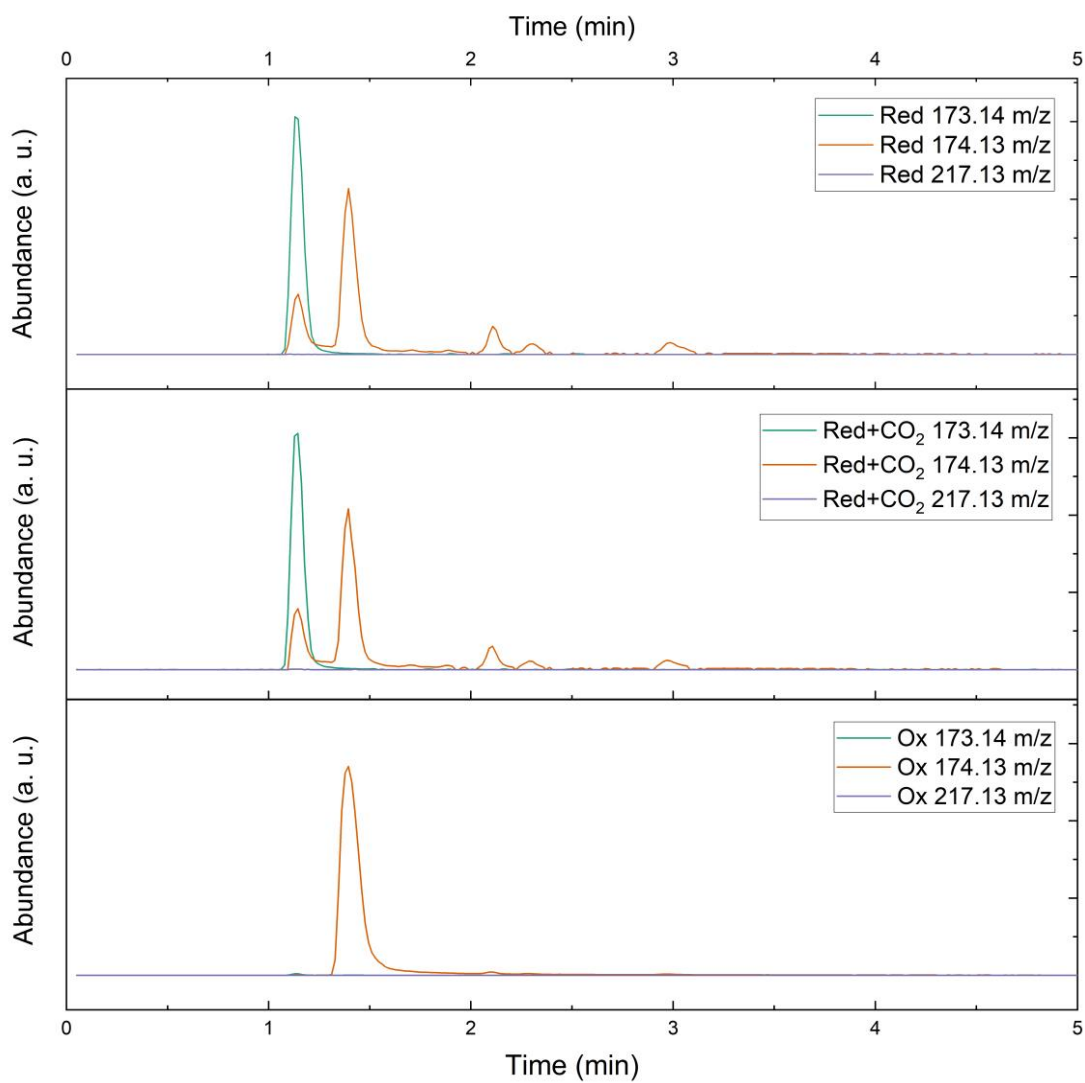

**Supplementary Fig. 11.** Extracted-ion chromatograms for the reduced state, the reduced state with $CO_2$ injected and the oxidized state of 50 μM H-TEMPO samples for 173.14, 174.13 and 217.13 m/z ions. 173.14 and 174.13 m/z ratios represent the parent H-TEMPO ion, where the difference in mass of ~ 1 Dalton is attributed to the addition of a hydrogen atom. The 217.13 m/z ratio represents the only identifiable peak that is close to a H-TEMPO-$CO_2$ adduct. 217.13 m/z ion has negligible abundance, signaling the absence of H-TEMPO-$CO_2$ adduct. This finding is also consistent with the proposed $CO_2$ capture mechanism of TEMPO, which is through bicarbonate formation rather than direct sorption onto TEMPO. At high pHs, the $OH^-$ combines with $CO_2$ to form bicarbonates while at low pHs, bicarbonate combines with $H^+$ to form carbonic acid that dissociates into $H_2O$ and $CO_2$ that is released. Note that a diluted concentration of 50 μM was used because of saturation limits of the spectrometer. In order to identify the origin of smaller 174.13 m/z peaks at times longer than 2 minutes, additional tandem MS/MS measurements would need to be done.

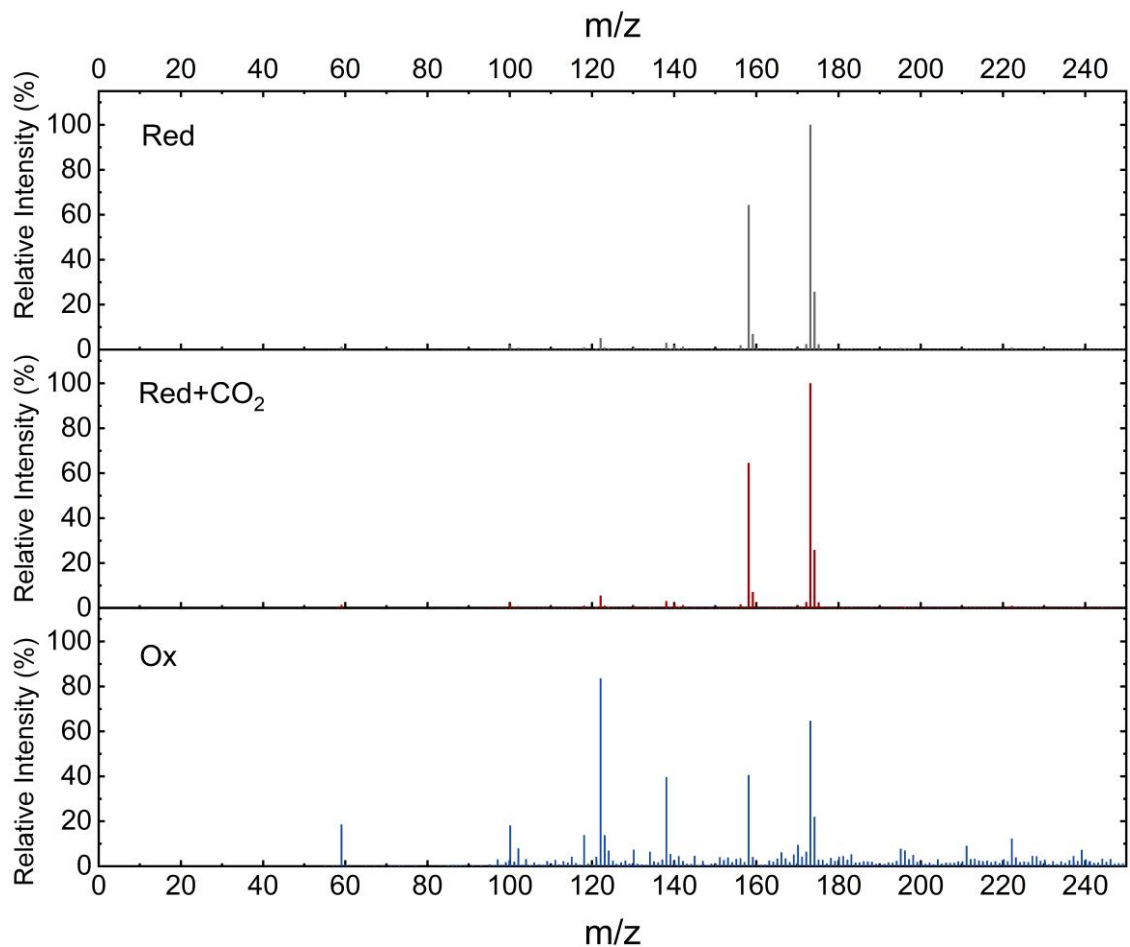

**Supplementary Fig. 12.** Mass spectra for the reduced state, the reduced state with $CO_2$ injected and the oxidized state of 50 μM H-TEMPO samples for a retention time of 1.14 min, showing the absence of signal at higher m/z than the parent peaks after $CO_2$ bubbling. For Red and Red+$CO_2$ samples, peaks at 173.14 and 174.13 m/z ratios represent the parent H-TEMPO, where the difference of ~ 1 Dalton can be attributed to an addition of a hydrogen atom. Peaks at 158.11 and 159.11 m/z are 15 Daltons smaller than the parent H-TEMPO, which represent the molecule after fragmentation of a methyl group with 15 Daltons. The fragmented methyl group is not shown because m/z ratios smaller than 30 were not collected because of the abundance of low mass molecules such as water. Elution of the oxidized sample at 1.14 min is small, as can be seen in Supplementary Fig. 11, and the spectra is noisier than Red and Red+$CO_2$. The mass spectra for Red and Red+$CO_2$ are qualitatively identical, suggesting that the molecular structure of H-TEMPO does not change with $CO_2$ injection and that $CO_2$ does not directly bind to TEMPO.

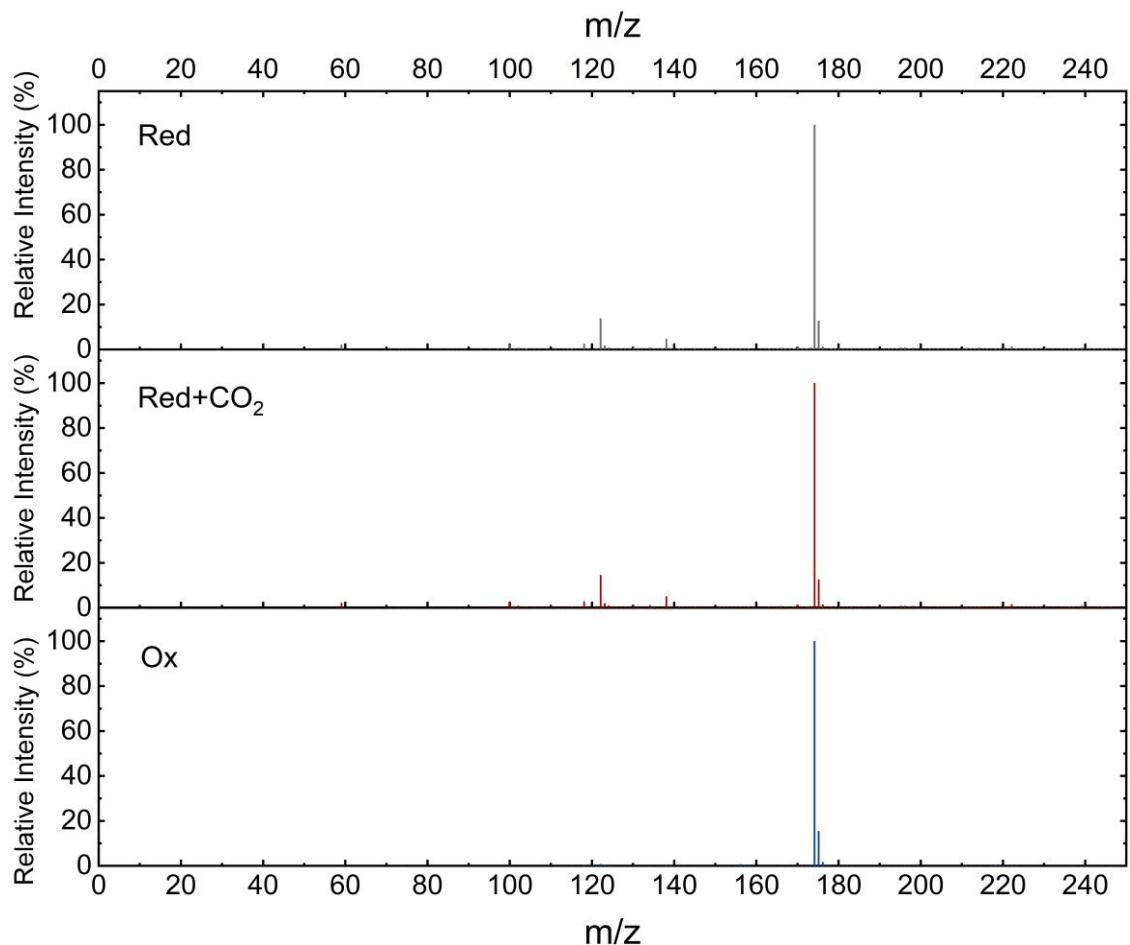

**Supplementary Fig. 13.** Mass spectra for the reduced state, the reduced state with $CO_2$ injected and the oxidized state of 50 μM H-TEMPO samples for a retention time of 1.39 min. The spectra show a predominant presence of the 174.13 m/z ratio from the parent H-TEMPO. Importantly, peaks at higher m/z ratios do not emerge for Red+$CO_2$. This result suggests that the direct sorption of $CO_2$ onto H-TEMPO does not take place.

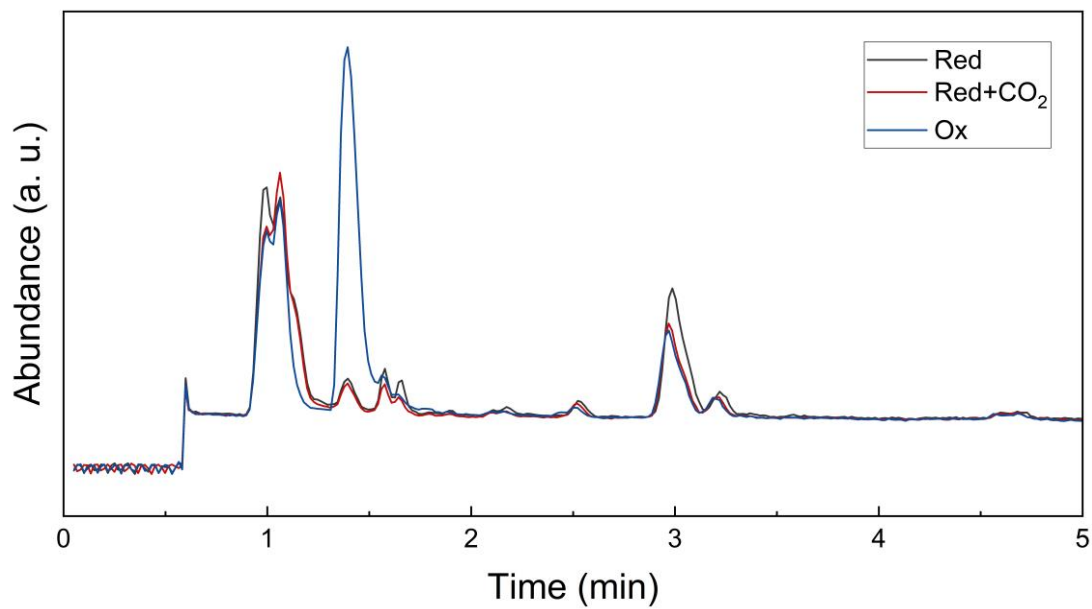

**Supplementary Fig. 14.** Total ion chromatograms for the reduced state, the reduced state with $CO_2$ injected and the oxidized state of 50 μM H-TEMPO samples. The major peaks at 1.14 min and 1.39 min correspond to H-TEMPO peaks, displayed in Supplementary Figs. 12-13, and peaks at around 1 min and 3 min are species present in the mobile phase.

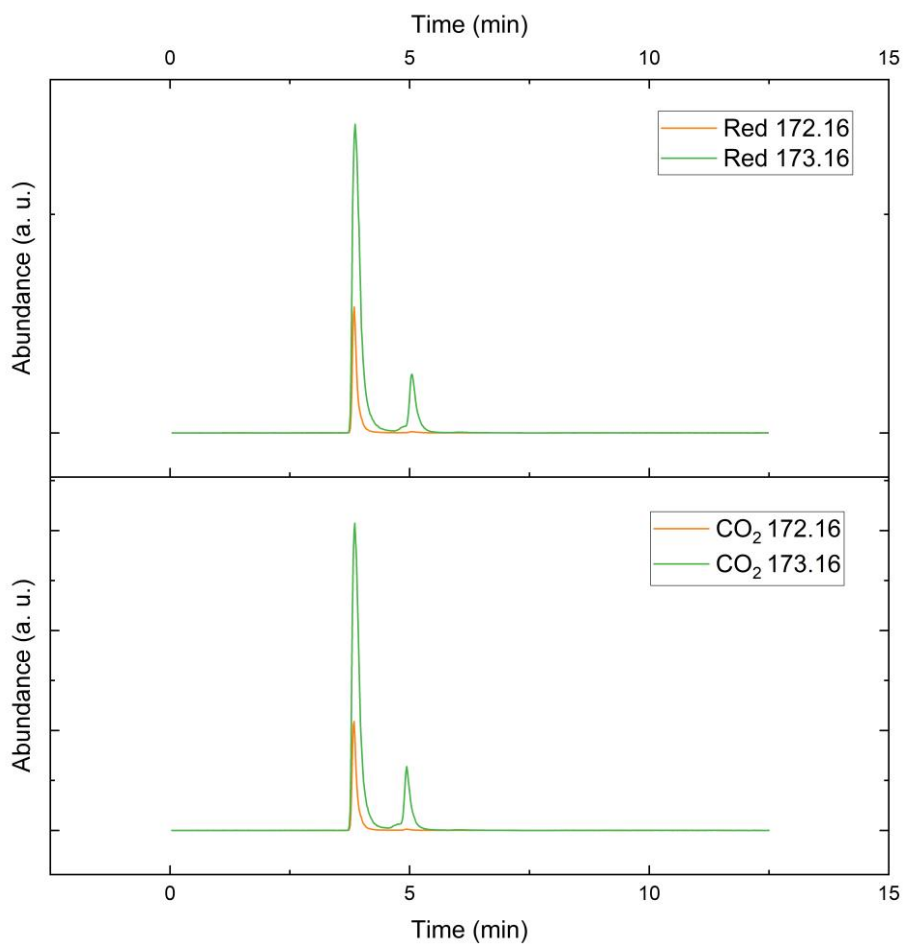

**Supplementary Fig. 15.** Extracted-ion chromatograms for the reduced and $CO_2$ bubbled 50 μM A-TEMPO samples for 172.16 and 173.16 m/z ions, both of which represent the parent A-TEMPO. The difference of 1 Dalton between the two ions can be attributed to the addition of a hydrogen atom. The two chromatograms show that A-TEMPO does not change structure after $CO_2$ injection. This result is consistent with the H-TEMPO results that $CO_2$ does not directly bind to TEMPO, even though A-TEMPO has a basic amine functional group that may be susceptible to $CO_2$ adduction.

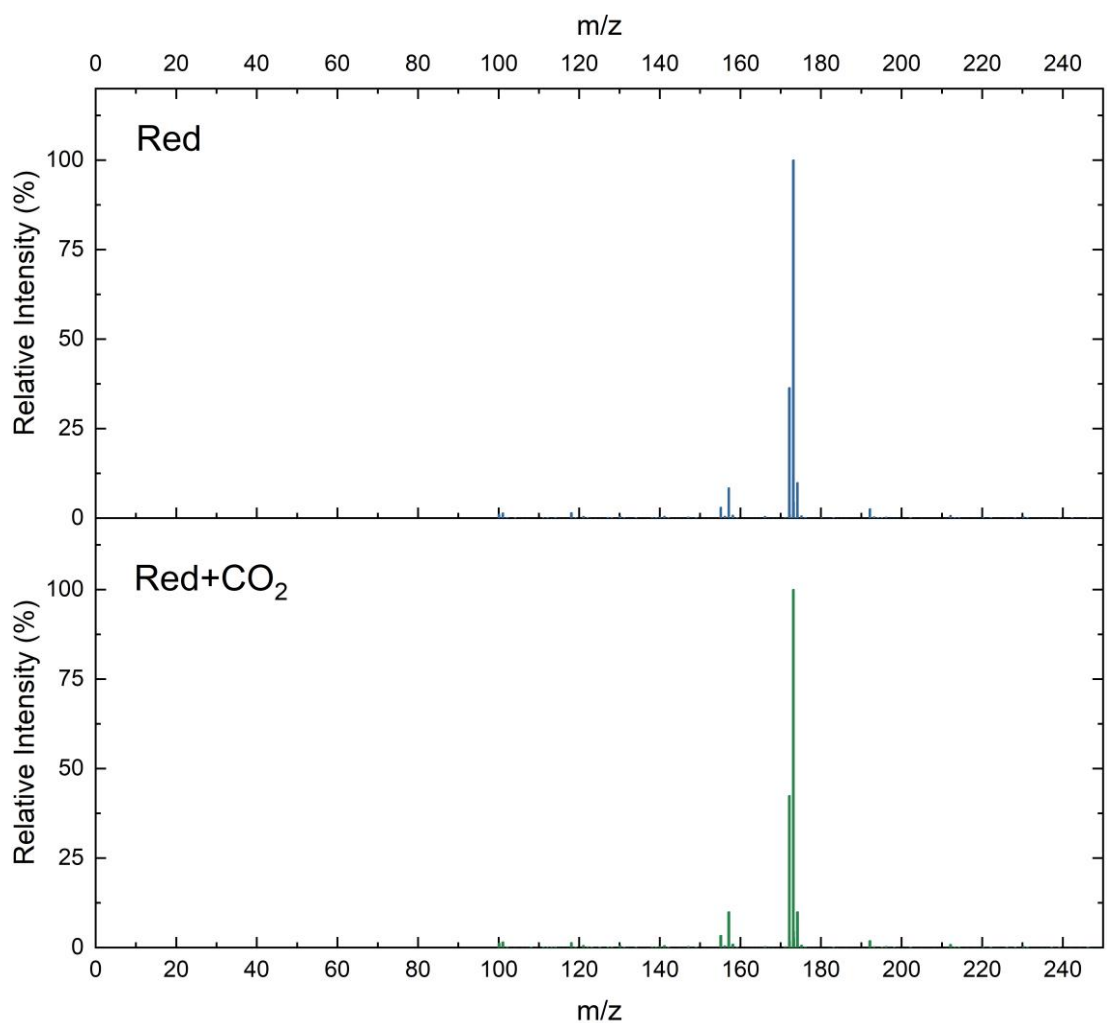

**Supplementary Fig. 16.** Mass spectra for reduced and $CO_2$ bubbled 50 μM A-TEMPO samples for a retention time of 3.84 min. The two most prominent peaks are at 173.16 and 172.16 m/z ratios, which correspond to the parent A-TEMPO. The 1 Dalton difference between the two ions can be attributed to the addition of a hydrogen atom. The spectra are qualitatively identical before and after $CO_2$ injection, suggesting that $CO_2$ does not directly bind to A-TEMPO.

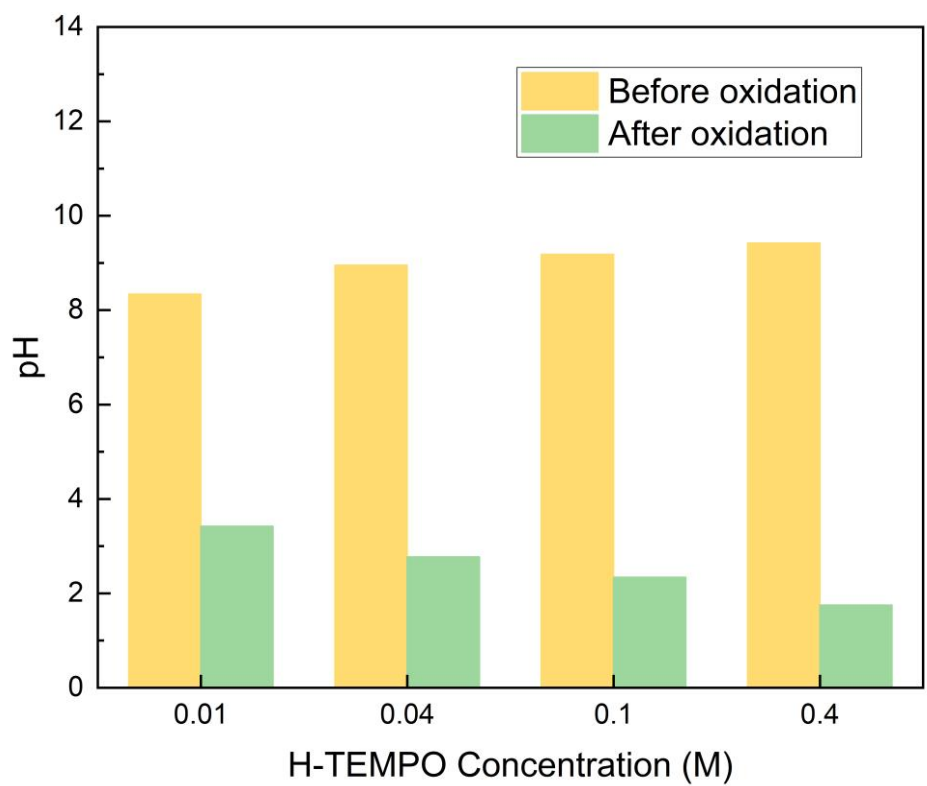

**Supplementary Fig. 17.** pH before and after oxidation for various concentrations of H-TEMPO solutions. Increasing H-TEMPO concentration leads to higher pH when reduced and lower pH when oxidized, allowing for larger pH swings at higher concentrations.

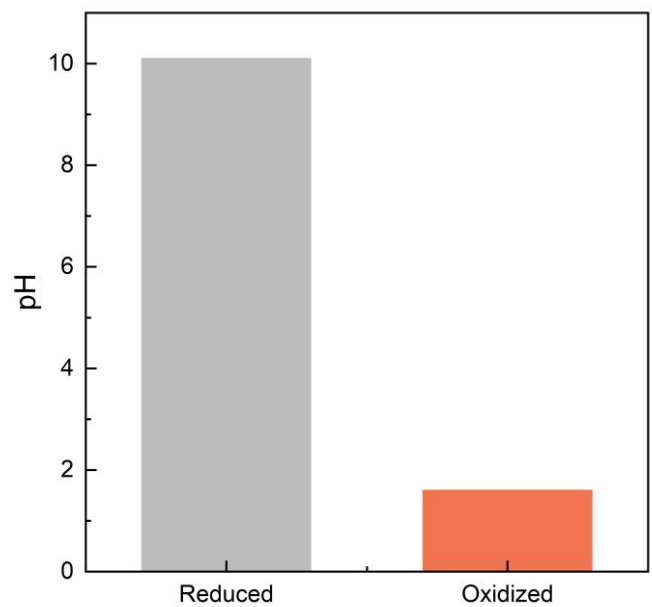

**Supplementary Fig. 18.** pH before and after oxidation for 100 mM A-TEMPO 400 mM KCl solution, showing that A-TEMPO oxidation allows for pH swings similar to H-TEMPO.

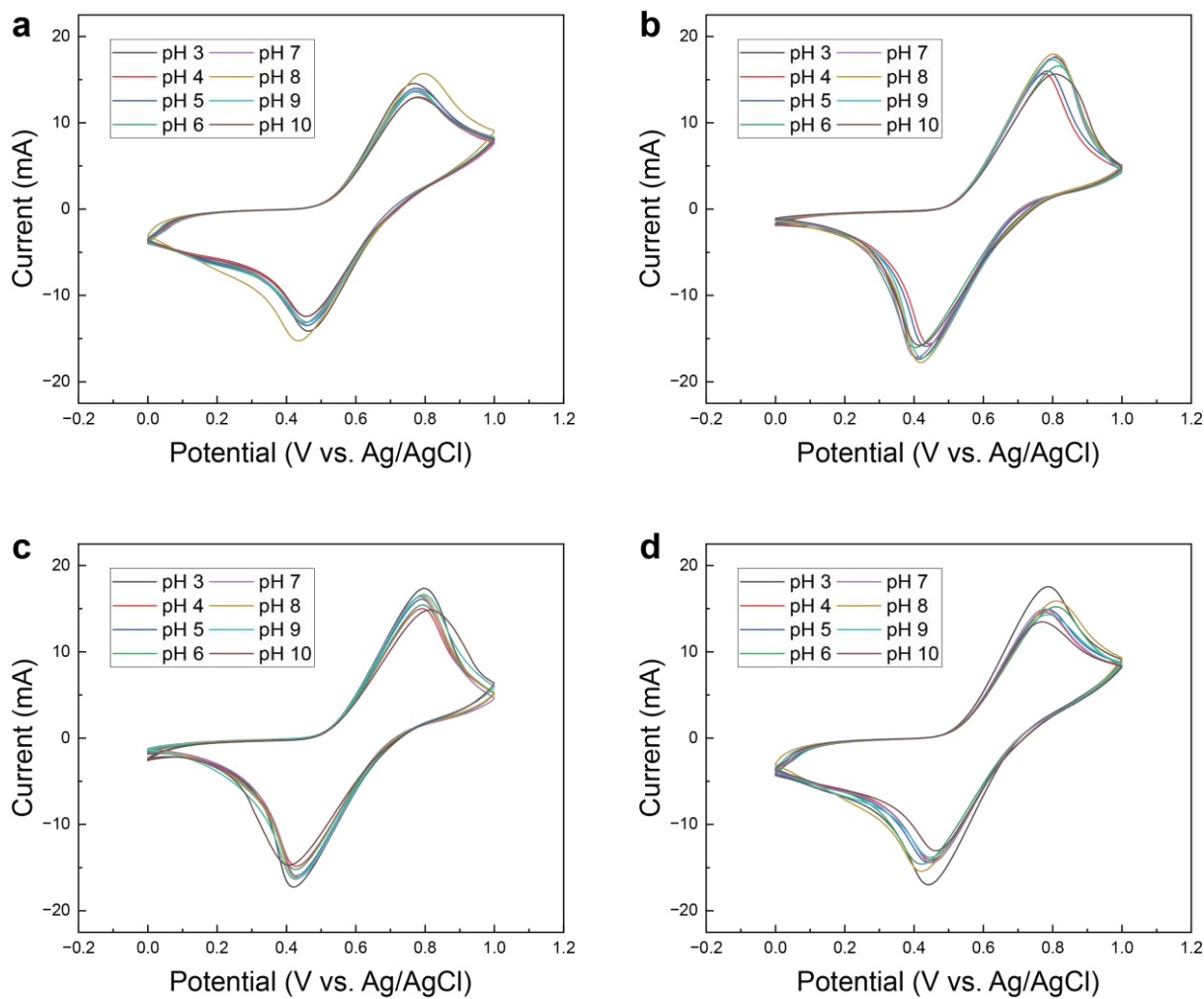

**Supplementary Fig. 19.** Cyclic voltammetry of 5 mM H-TEMPO 100 mM KCl electrolyte at different pHs. A total of 5 sets of data were collected, where four are shown here and the fifth is shown in the inset of Fig. 3b. Using the midpoint of the oxidative and reductive peaks, the equilibrium potential plot shown in Fig. 3b was constructed. The slight differences in the shape of the CV curves are products of the small differences in experimental conditions, such as the graphite electrode size and the electrode immersion depth that impacts the active surface area. However, the electrochemical properties are qualitatively independent of the pH, and the midpoint is quantitatively independent of pH in these pH ranges.

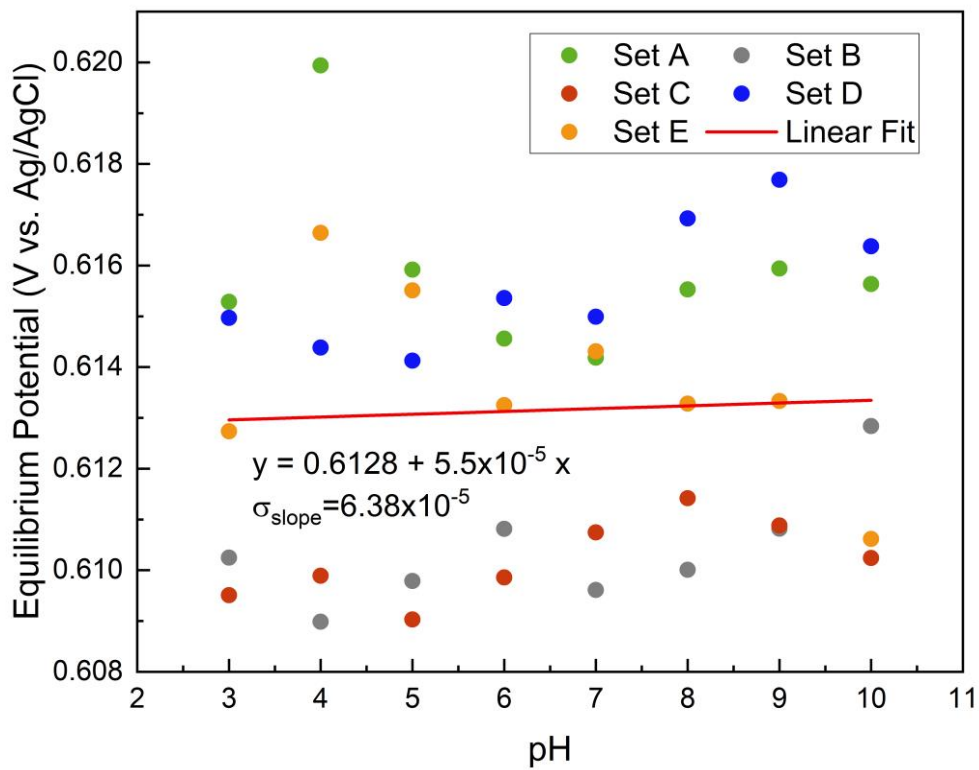

**Supplementary Fig. 20.** Equilibrium potential data acquired from the cyclic voltammetry results shown in Figs. 3b and Supplementary Fig. 19. Set A-D correspond to panels a-d in Supplementary Fig. 19 and Set E is shown in Fig. 3b. The time-ordering of the pH values for each data set was chosen at random. A linear fit of the data has a slope of $5.5 \times 10^{-5}$ V/pH, with a standard deviation of $6.38 \times 10^{-5}$ V/pH. The standard deviation was obtained by combining the standard deviations of each set. The lower bound of the slope, obtained through slope minus σ, is $-8.8 \times 10^{-6}$ V/pH. In contrast, conventional pH-dependent redox chemistries have a dependence of $-5.9 \times 10^{-2}$ V/pH, which is 925 σ away from our results, illustrating the statistical difference in the pH-dependence between conventional pH swing methods and our chemistry.

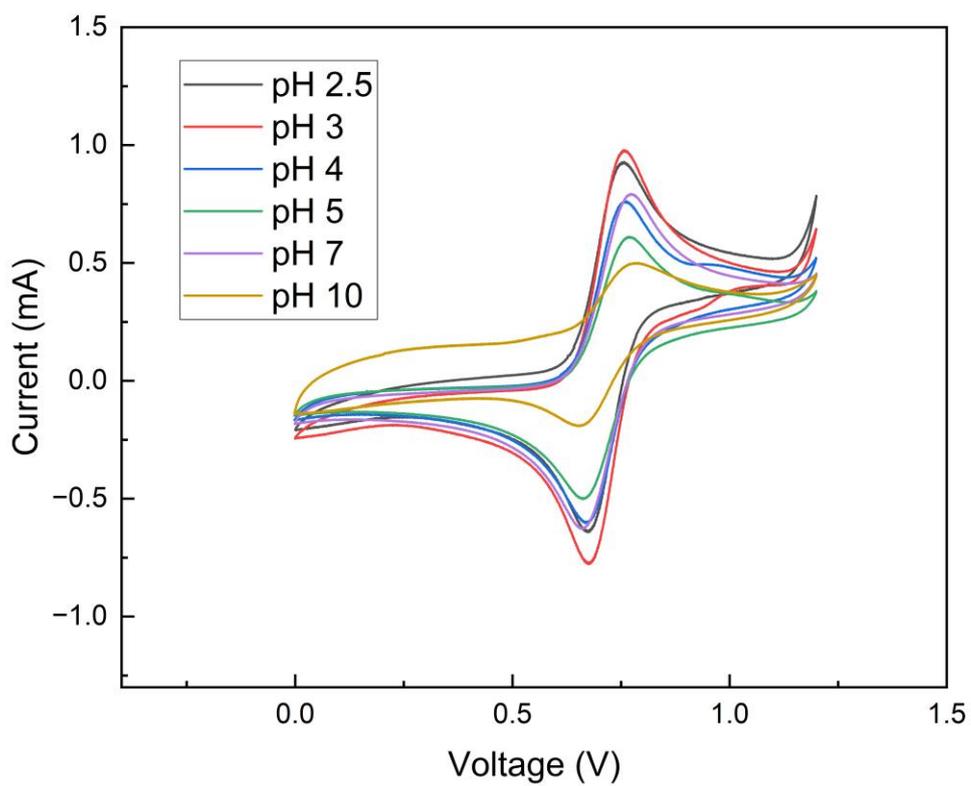

**Supplementary Fig. 21.** Cyclic voltammetry of 5 mM A-TEMPO 100 mM KCl at various pHs. Similar to H-TEMPO, A-TEMPO also has a pH-independent equilibrium potential, which can be obtained by the midpoint of the reductive and oxidative peaks. The midpoint vs. pH is plotted in Supplementary Fig. 22.

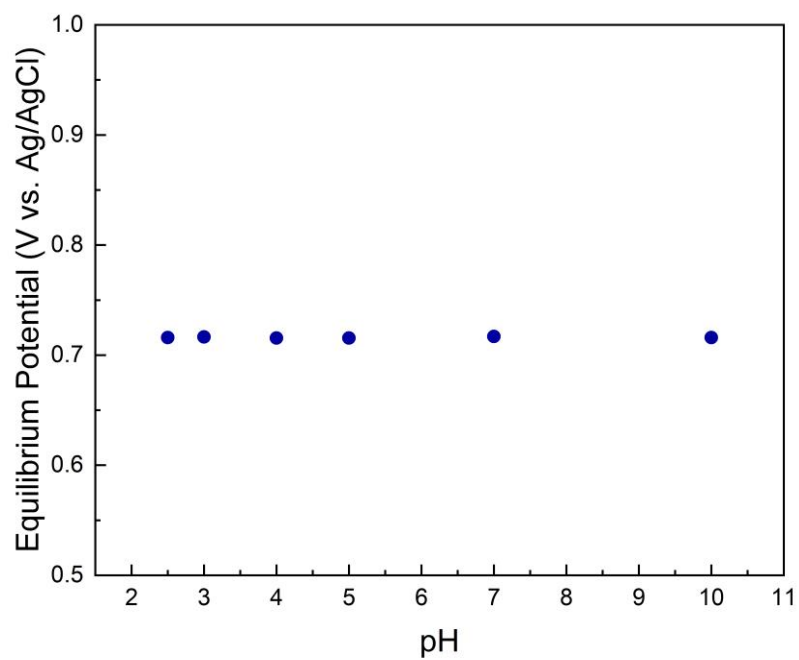

**Supplementary Fig. 22.** Equilibrium redox potential of 5 mM A-TEMPO 100 mM KCl at various pHs, obtained from the midpoint of the oxidative and reductive peaks in cyclic voltammetry plots in Supplementary Fig. 21.

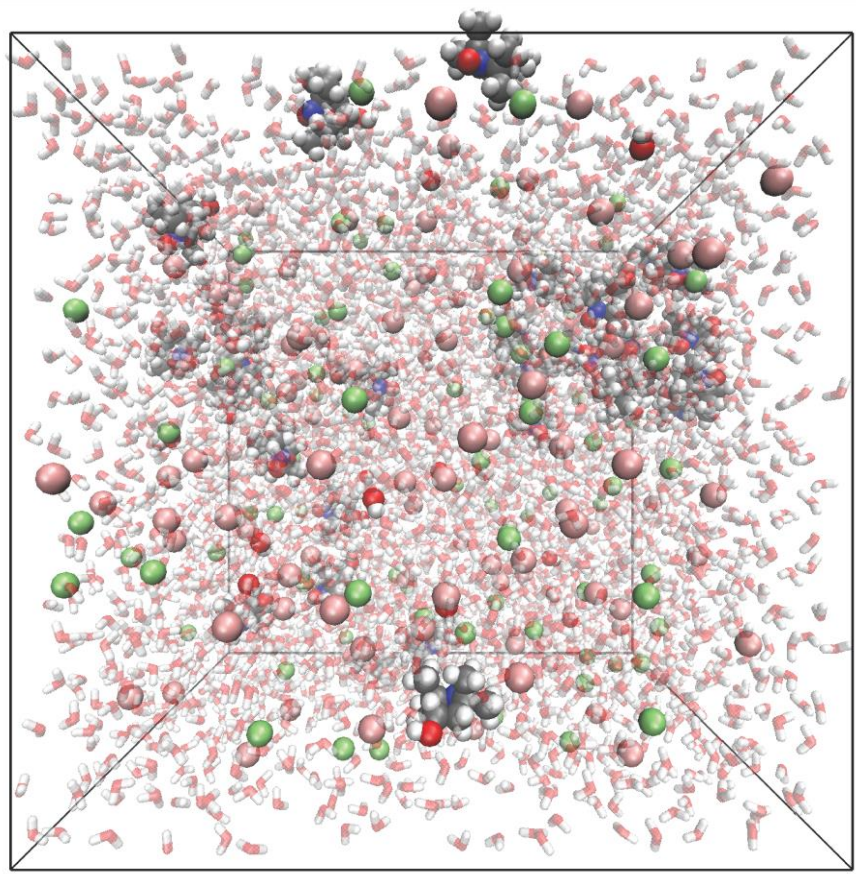
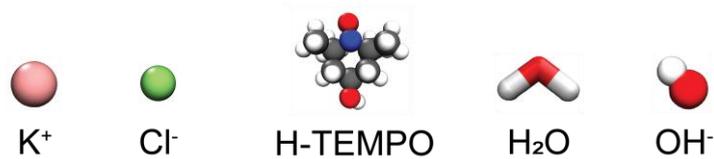

**Supplementary Fig. 23.** Molecular dynamics simulations of 0.4 M reduced H-TEMPO with 1.2 M KCl 0.2 M KOH electrolyte.

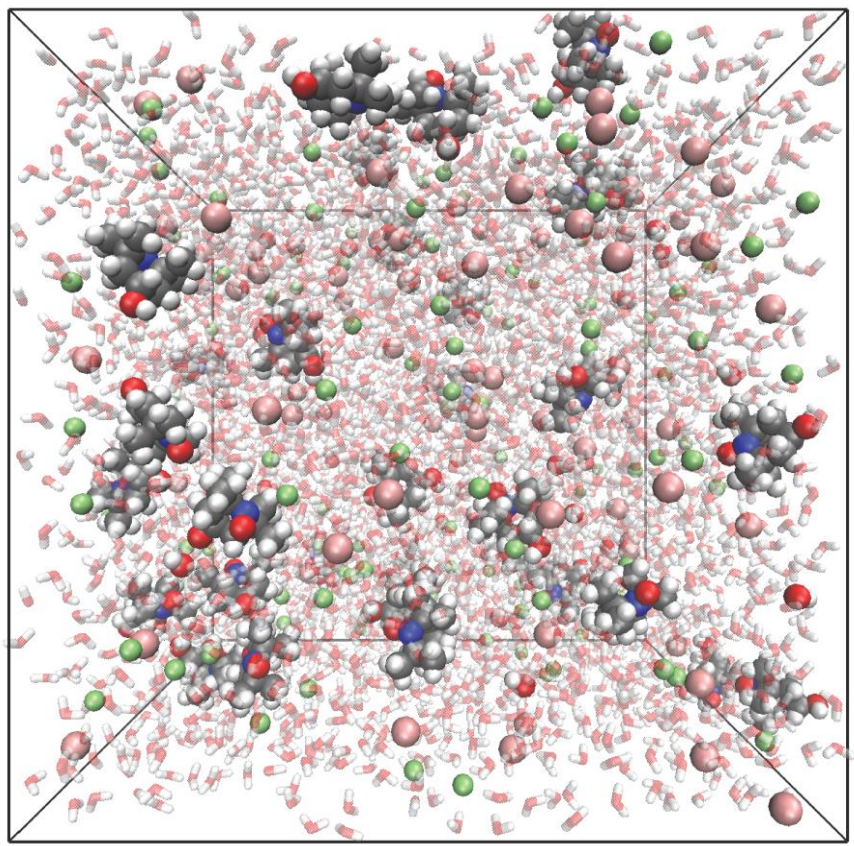

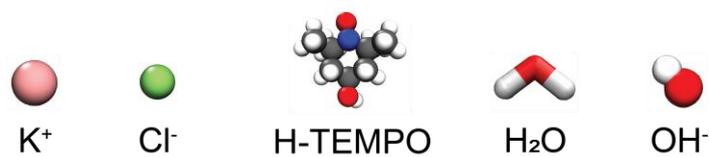

**Supplementary Fig. 24.** Molecular dynamics simulations of 0.4 M oxidized H-TEMPO with 1.2 M $K^+$ 1.6 M $Cl^-$ 0.2 M KOH electrolyte.

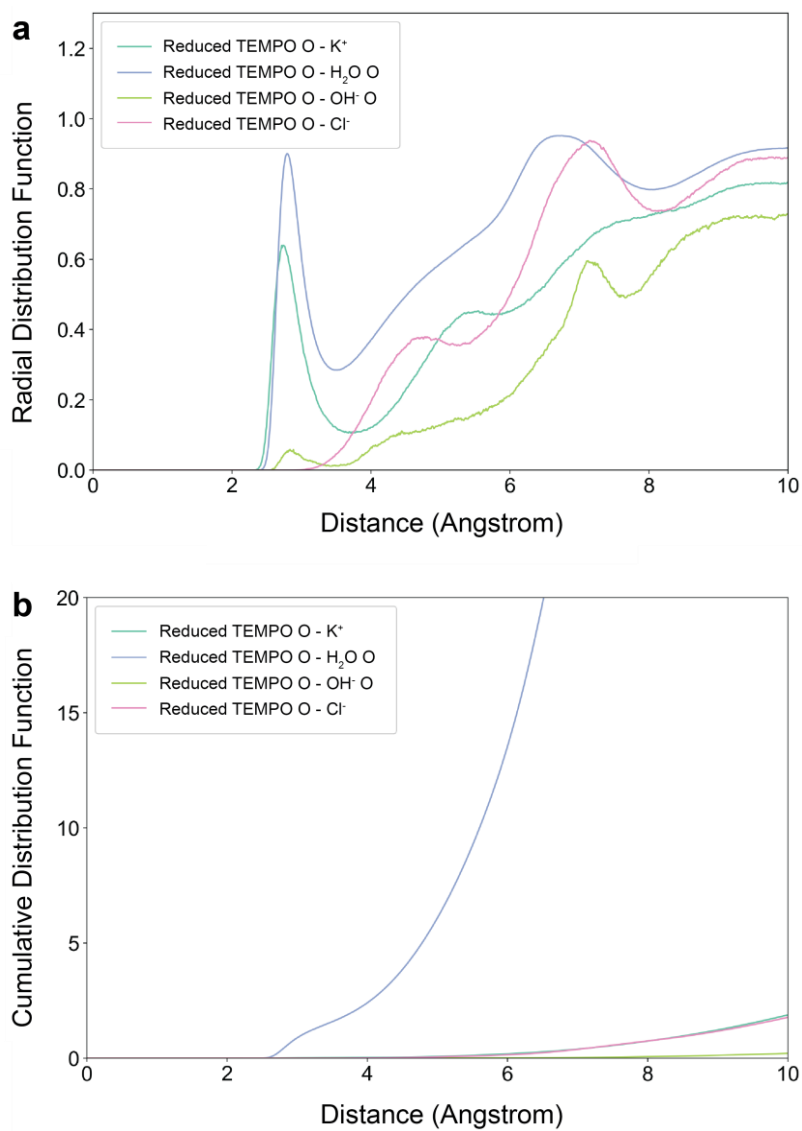

**Supplementary Fig. 25.** Molecular dynamics simulations of **a**, the radial distribution functions, $g(r) = \langle \rho(r) \rangle / \rho_{avg}$ where $\rho$ is the density of the oxygen atom on species $K^+$ and the oxygen atom attached to water (abbreviated $H_2O$ O) around the nitroxyl O on H-TEMPO and **b,** the cumulative distribution functions, defined as $4\pi \rho_{avg} \int r^2 g(r) dr$, for the nitroxyl O on H-TEMPO in 0.4 M reduced H-TEMPO with 1.2 M KCl 0.2 M KOH electrolyte. Several insights can be deduced from these plots. First, it is not favorable for $OH^-$ and $Cl^-$ to associate with H-TEMPO oxygen, while it is favorable for water and $K^+$ to coordinate with H-TEMPO oxygen. Second, due to the large number of water molecules (~55 mol/L), the coordination number around H-TEMPO oxygen is dominated by water. Third, the radial distance of water O depends on the oxidation state, which can be seen by comparing TEMPO O – $H_2O$ O with the oxidized counterpart in Supplementary Fig. 27.

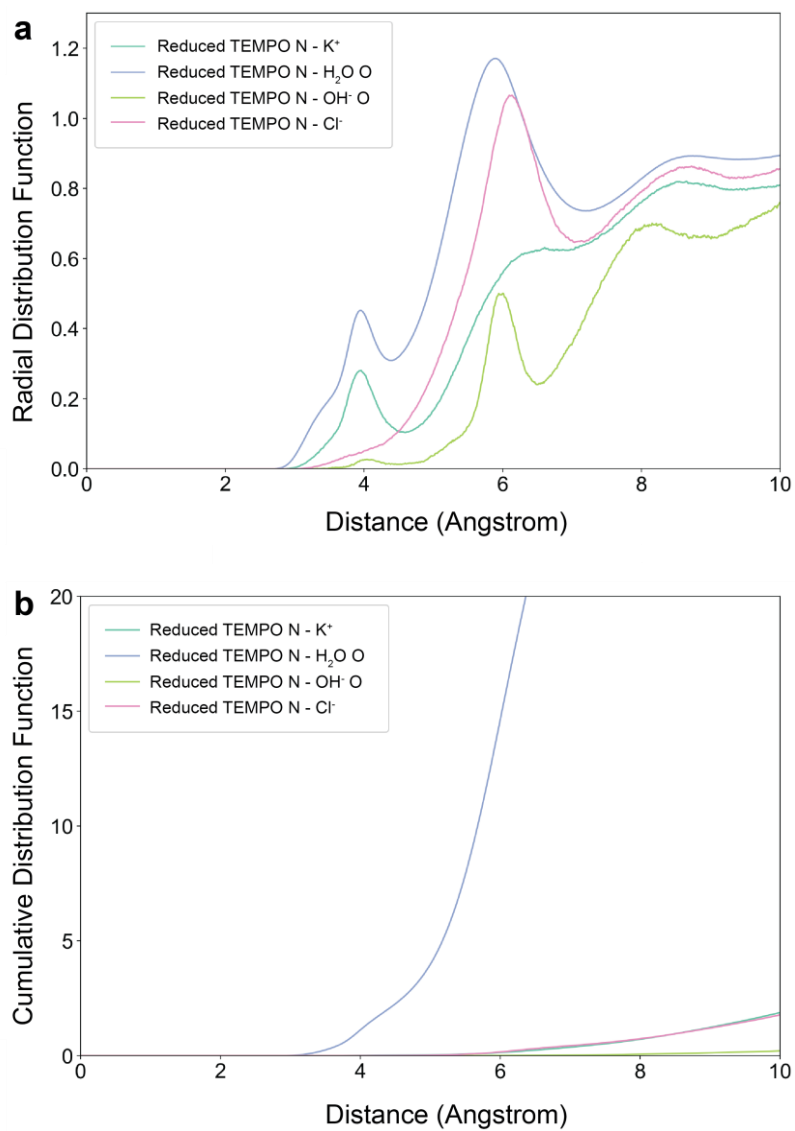

**Supplementary Fig. 26.** Molecular dynamics simulations of **a,** the radial distribution functions and **b,** the cumulative distribution functions, defined as $4\pi\rho_{avg}\int r^2 g(r)dr$, which are essentially the coordination numbers for the nitroxyl N on H-TEMPO in 0.4 M reduced H-TEMPO with 1.2 M KCl 0.2 M KOH electrolyte. Similar to oxygen, several insights can be deduced from these plots. First, it is not favorable for $OH^-$ and $Cl^-$ to associate with H-TEMPO nitrogen, while it is favorable for water and $K^+$ to coordinate with H-TEMPO nitrogen. Second, due to the large number of water molecules (~55 mol/L), the coordination number around H-TEMPO nitrogen is dominated by water. Third, the radial distance of water O depends on the oxidation state, which can be seen by comparing TEMPO N – $H_2O$ O with the oxidized counterpart in Supplementary Fig. 28.

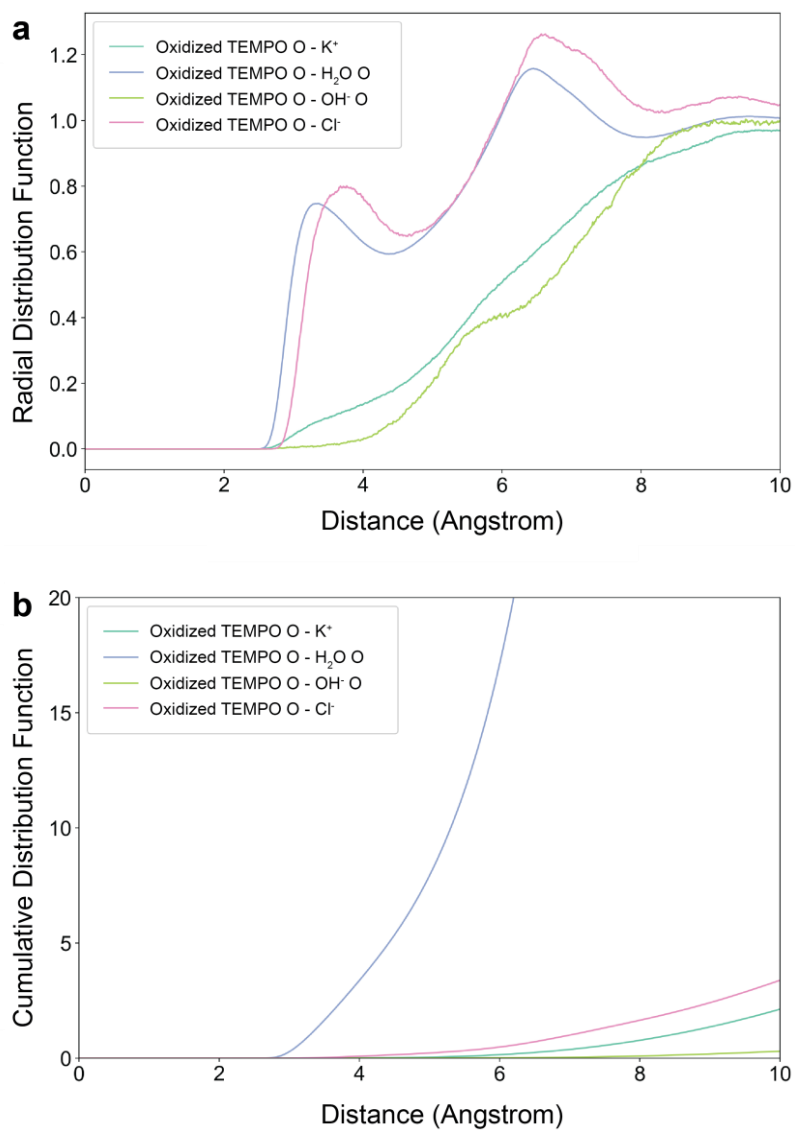

**Supplementary Fig. 27.** Molecular dynamics simulations of **a,** the radial distribution functions and **b,** the cumulative distribution functions, which are essentially the coordination numbers for the nitroxyl O on H-TEMPO in 0.4 M oxidized H-TEMPO with 1.2 M $K^+$ 1.6 M $Cl^-$ 0.2 M KOH electrolyte. Unlike the reduced counterpart, oxidized H-TEMPO prefers to interact with water and $Cl^-$, while $OH^-$ and $K^+$ are located relatively farther away. It is important to note that $OH^-$ does not coordinate with the positively charged H-TEMPO. Similar to reduced TEMPO, hydration layer is dominated by water molecules due to the large number.

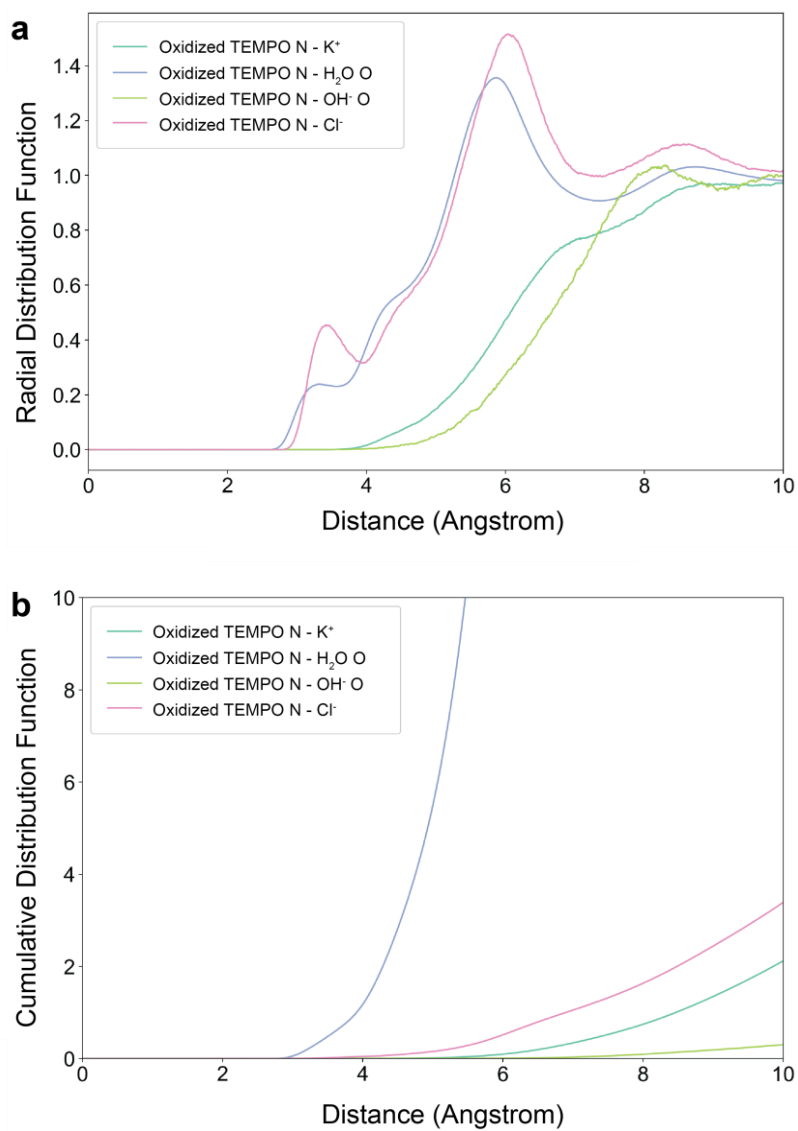

**Supplementary Fig. 28.** Molecular dynamics simulations of **a,** the radial distribution functions and **b,** the coordination function between the oxidized TEMPO of the oxygen atoms in water, which are essentially the coordination numbers for the nitroxyl N on H-TEMPO in 0.4 M oxidized H-TEMPO with 1.2 M $K^+$ 1.6 M $Cl^-$ 0.2 M KOH electrolyte.

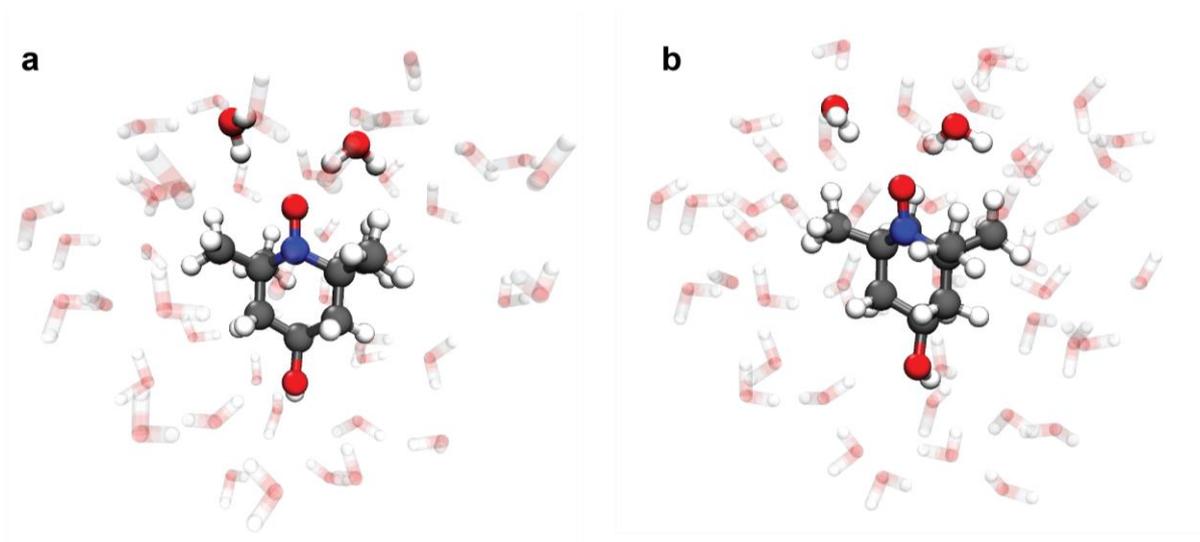

**Supplementary Fig. 29.** Snapshots of reduced H-TEMPO and water interactions simulated with molecular dynamics, showing that the nitroxyl O interacts closely with water H. This is due to the negative partial charge on the nitroxyl O, as shown by DFT simulations in Fig. 3f, polarizing and orienting water H. This has important implications on the activity coefficient ($\gamma$) of $H^+$ and pH, defined as $pH \equiv -\log_{10}(\gamma[H^+])$. Due to the partial negative charge on the nitroxyl functional group of TEMPO, water molecules are also negatively polarized, as shown by Fig. 3g. This can decrease the activity coefficient of surrounding $H^+$ by stabilizing it through electrostatic interactions. In addition, it can suppress the formation of $H^+$ in the presence of other species such as $CO_2$ or bicarbonates. This causes the low activity coefficient and high pH of reduced H-TEMPO.

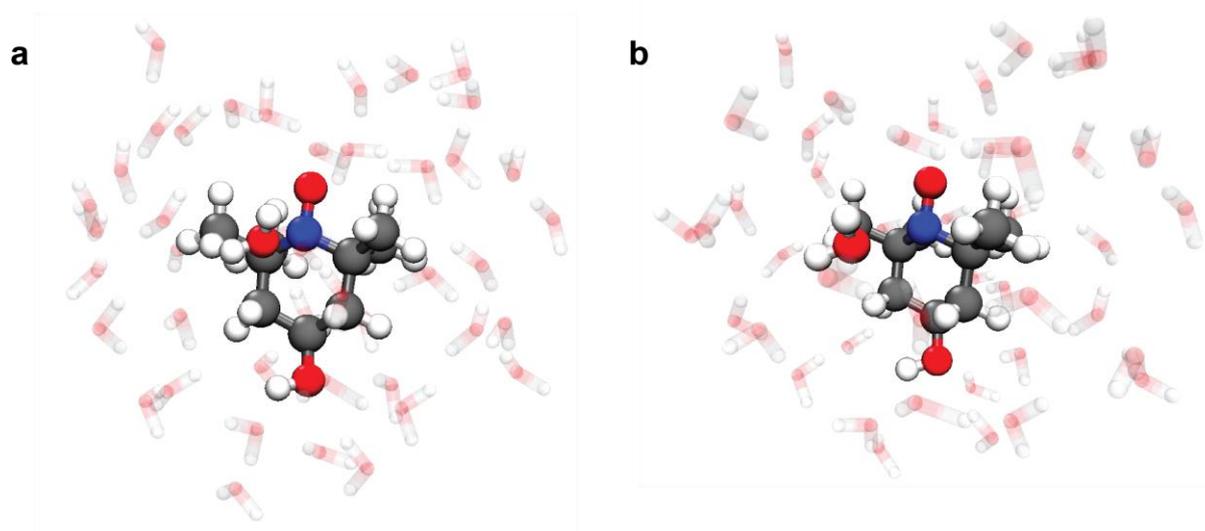

**Supplementary Fig. 30.** Snapshots of oxidized H-TEMPO and water interactions simulated with molecular dynamics. They show that the nitroxyl N interacts closely with water O, owing to the positive partial charge on the nitroxyl N, as shown in Fig. 3f. This interaction positively polarizes the surrounding water molecules. Positively polarized water has a profound impact on the activity coefficient ($\gamma$) of H$^+$ and pH, defined as $pH \equiv -\log_{10}(\gamma[H^+])$. Importantly, TEMPO oxidation modulates the pH by changing the activity coefficient of H$^+$ through water polarization rather than direct formation of H$^+$ ions.

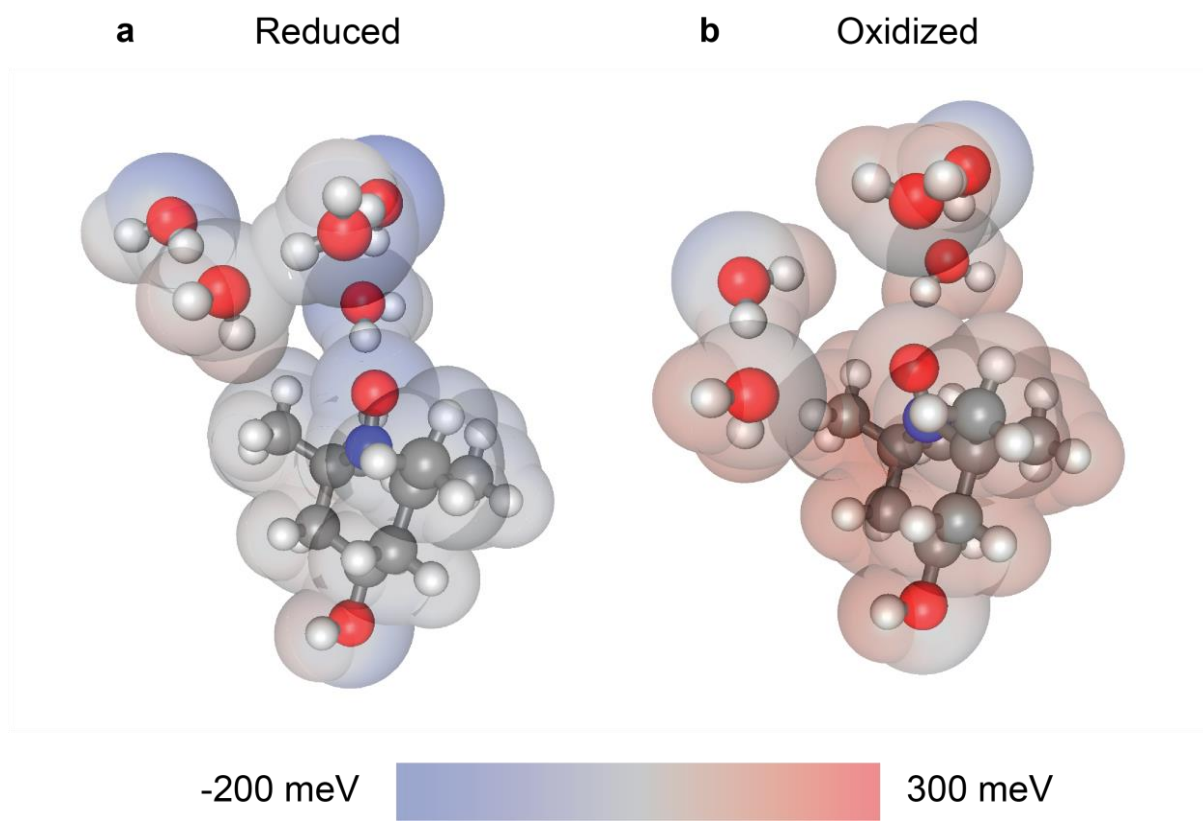

**Supplementary Fig. 31.** Electrostatic potential surfaces of **a,** reduced and **b,** oxidized H-TEMPO and surrounding water simulated with density functional theory. In **a**, the surrounding water molecules are negatively polarized whereas in **b**, they are more positively polarized. This has important implications on the activity of $H^+$ in the solution. The negatively polarized water molecules stabilize $H^+$ through electrostatic interactions, suppressing the formation of additional $H^+$. On the contrary, positively polarized water molecules destabilize $H^+$ and provide additional $H^+$ in the presence of an electron withdrawing and proton-accepting species such as bicarbonates. These effects on water profoundly affect the activity coefficient of $H^+$. Therefore, by oxidizing TEMPO, large swings in pH can be achieved without directly modulating the $H^+$ concentration.

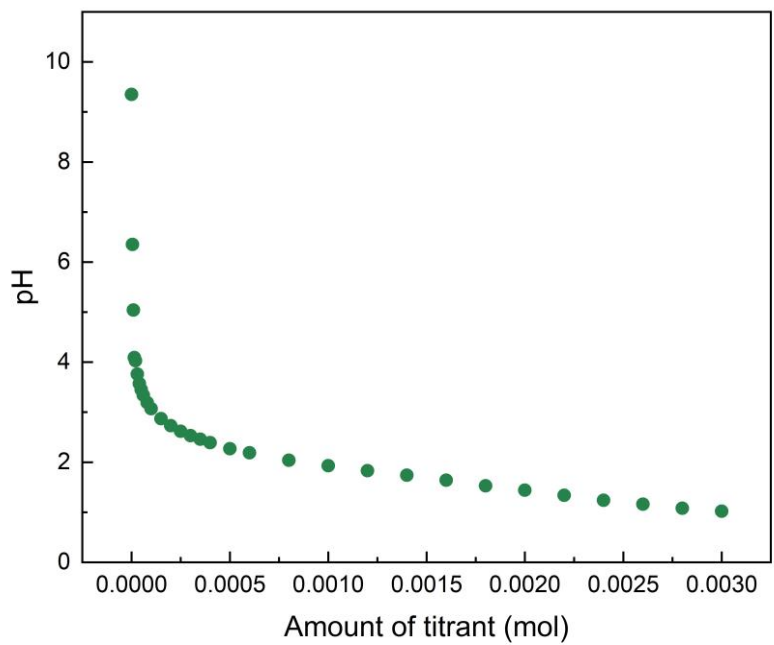

**Supplementary Fig. 32.** Titration of 10 mL of 0.4 M reduced H-TEMPO in 1.2 M KCl solution with 1 M HCl titrant. The reduced H-TEMPO solution initially had a pH of 9.4. The pH drops precipitously with addition of HCl titrant. After the initial steep decrease to pH ~ 2, the pH decreases with a moderate slope towards that of the titrant. This result is in agreement with the observation with Fig. 3a, where we see a steep decrease in pH as the reduced H-TEMPO is oxidized into TEMPO$^+$, an electron acceptor Lewis acid.

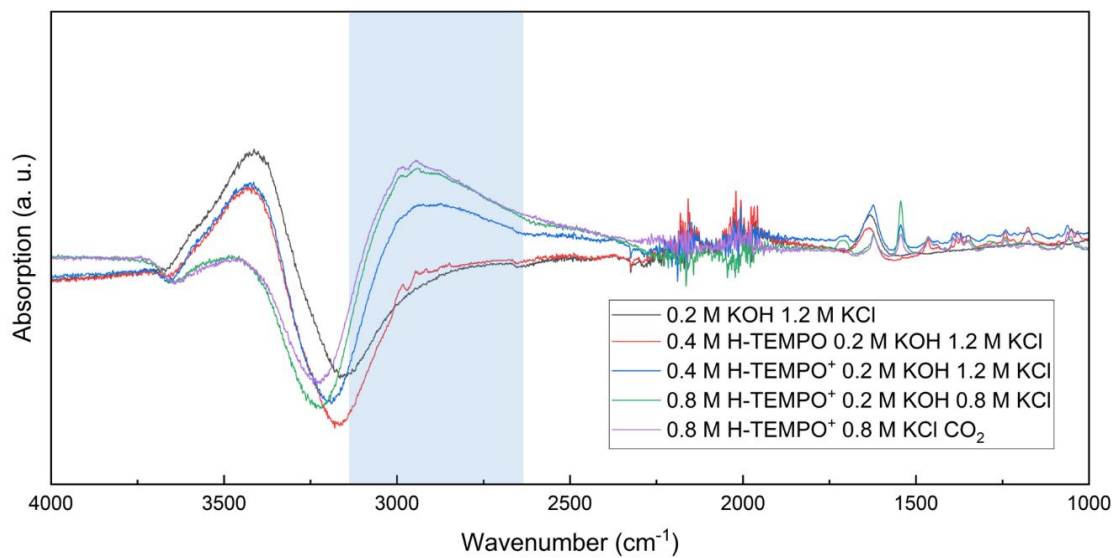

**Supplementary Fig. 33.** *Ex-situ* FT-IR spectra of various H-TEMPO solutions, all obtained with water background subtraction. A broad peak around 2900 cm$^{-1}$ emerges for solutions with H-TEMPO$^+$, which is hypothesized to represent the appearance of weak interactions between H-TEMPO$^+$ and H$_2$O. At 0.8 M H-TEMPO$^+$ concentration, the peak is larger than compared to 0.4 M H-TEMPO$^+$, consistent with the hypothesis. However, we are unable identify the vibrational mode that would be modified by the TEMPO$^+$ - water interactions. The peak from 3400-3500 cm$^{-1}$ is believed to be O-H stretch due to KCl – H$_2$O interactions, and the trough around 3200 cm$^{-1}$ may be from water displacement. An unambiguous signature of an increase of the OH stretch vibration is hidden in the vibrational spectra of the much more abundant water molecules. As water has strong absorption signal from 2800 – 3700 cm$^{-1}$, displacement of water by addition of other species such as H-TEMPO and KCl can result in a negative signal. The peaks from 1000 – 1600 cm$^{-1}$ correspond to TEMPO and TEMPO$^+$, as they are absent in 0.2 M KOH 1.2 M KCl.

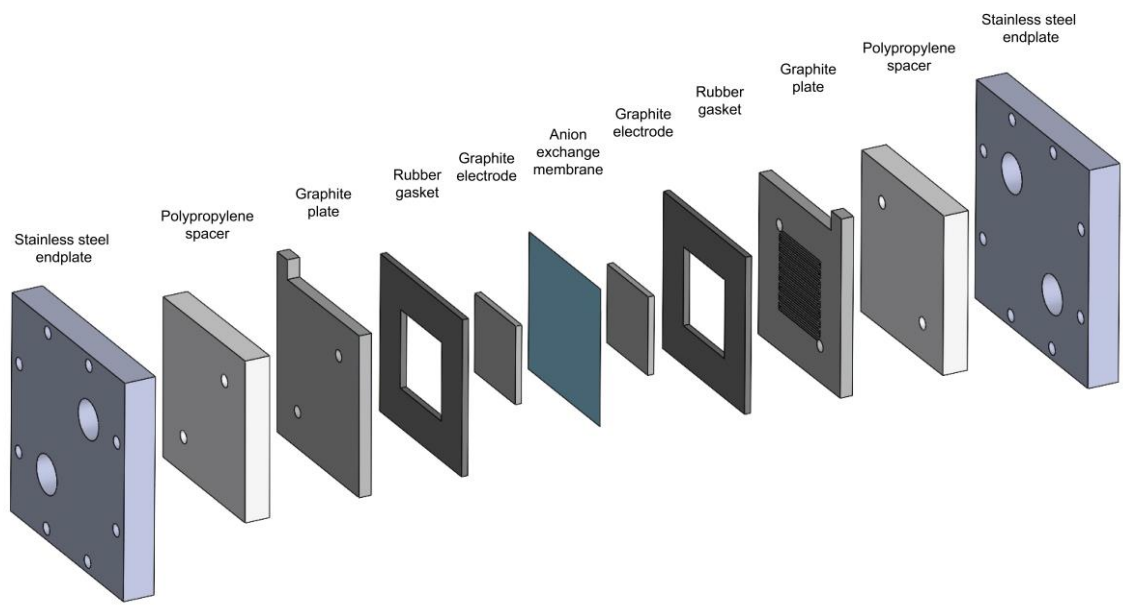

**Supplementary Fig. 34.** Exploded view drawing of the electrochemical cell design, composed of a graphite felt electrode, gasket, flow-patterned graphite plate, polypropylene spacer and stainless-steel plate. The overall resistance of this electrochemical can be significantly reduced. A ~ 2-fold lower anion membrane has been tested. The thickness of graphite felt electrode (3.1 mm in this work) can be reduced to < 0.5 mm, and the dimensions of the serpentine path in the graphite plates (1.25 mm width x 1.25 mm depth) will be optimized to further reduce the ionic conductivity while maintaining low fluid flow resistance.

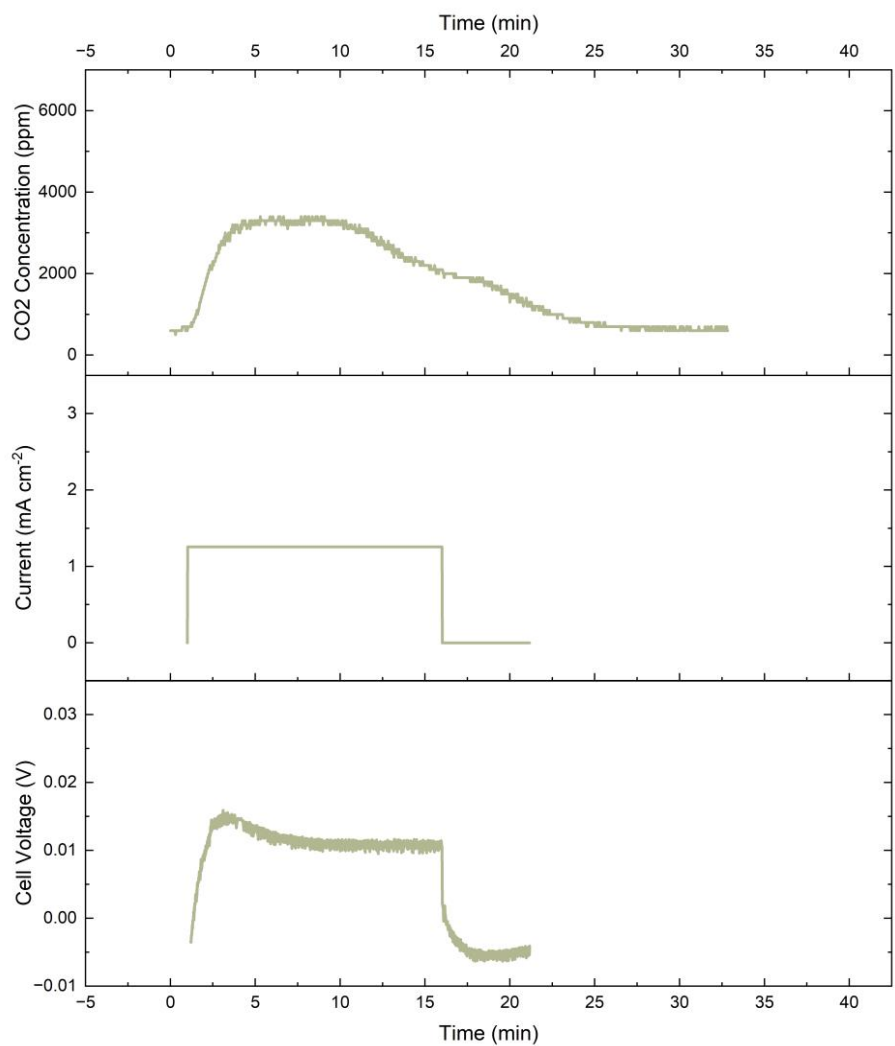

**Supplementary Fig. 35.** Flow cell $CO_2$ capture and release performance using 100 mM A-TEMPO 400 mM KCl solution. It can be seen that at a current density of 1.25 mA cm$^{-2}$, only a small voltage of ~ 0.01 V is needed for TEMPO redox and $CO_2$ capture/release.

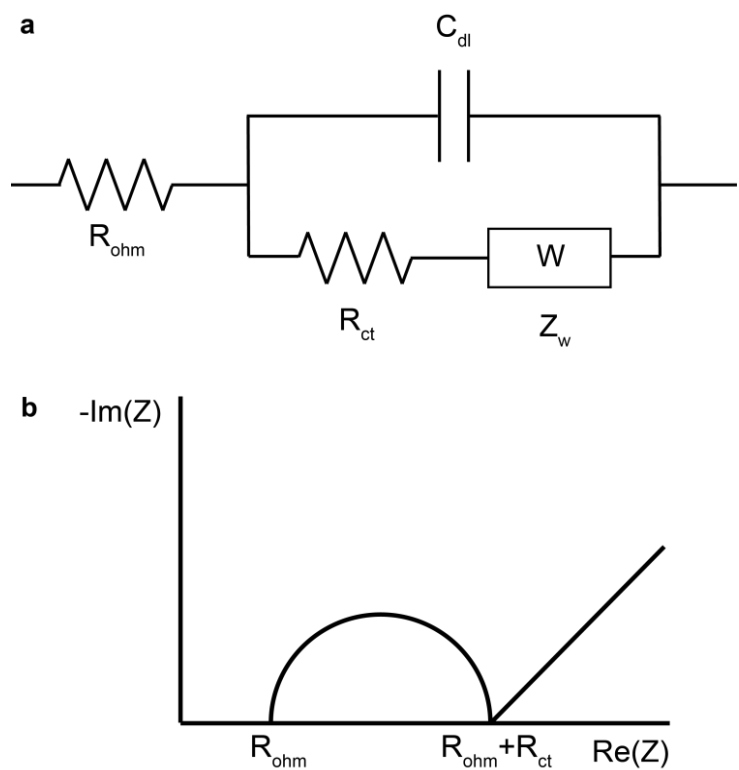

**Supplementary Fig. 36.** Electrochemical impedance spectroscopy of the electrochemical cell in the flow system. **a**, an idealized equivalent circuit model of the electrochemical cell, where $R_{ohm}$ is the ohmic resistance of electron and ion transport, $C_{dl}$ is the capacitance of the double layer, $R_{ct}$ is the charge-transfer resistance, and $Z_W$ is the Warburg impedance. **b**, Nyquist plot of the equivalent circuit model; detailed analysis of the results are discussed in Supplementary Note 4.